\documentclass[%
 reprint,
superscriptaddress,
amsmath,amssymb,
aps,
prb,
]{revtex4-2}

\usepackage{graphicx}
\usepackage{dcolumn}
\usepackage[english]{babel}
\usepackage{bm}
\usepackage{braket}
\usepackage{float}

\graphicspath{{Figures/}}

\newcommand{\tr}{$\mathcal{T}$ }

\usepackage{svg}
\usepackage{hyperref}
\usepackage{cleveref}
\usepackage{amssymb}
\usepackage{pifont}

\begin{document}

\title{Noncollinear Magnetic Multipoles in Collinear Altermagnets}

\author{Luca Buiarelli}
\affiliation{Department of Chemical Engineering and Materials Science, University of Minnesota, MN 55455, USA}

\author{Rafael M. Fernandes}
\affiliation{Department of Physics, The Grainger College of Engineering, University of Illinois Urbana-Champaign, Urbana, IL 61801, USA}
\affiliation{Anthony J. Leggett Institute for Condensed Matter Theory, The Grainger College of Engineering, University of Illinois Urbana-Champaign, Urbana, 61801, IL, USA}

\author{Turan Birol}
\thanks{Corresponding author: \href{mailto:tbirol@umn.edu}{tbirol@umn.edu}}
\affiliation{Department of Chemical Engineering and Materials Science, University of Minnesota, MN 55455, USA}

\date{\today}

\begin{abstract}
Altermagnets host an array of magnetic multipoles, which are often visualized and studied in the reciprocal space. In the real space, the relative phase of the multipoles of the spin-density around atoms determines whether a system is an altermagnet or a conventional antiferromagnet. In this study, we approach these real space multipoles in altermagnets using a combination of first principles calculations and group theory. We show that even in collinear magnets, the local spin density is necessarily noncollinear due to spin-orbit coupling. Moreover, the noncollinear contributions often provide a more direct illustration of the magnetic multipolar character of altermagnetism than the collinear contribution, which is dominated by the dipolar term. Our first principles calculations also show that 32-poles, in addition to the octupoles, can be visible in spin-density of d-wave altermagnets, and they must be taken into account in discussions of the macroscopic response. Finally, we elucidate the interplay between magnetism and subtle crystal structural distortions in perovskite altermagnets, which provide a fertile playground for studying phase transitions between antiferromagnetic and altermagnetic phases. 
\end{abstract}
\maketitle

\section{Introduction} 
Altermagnetism (AM) has recently been identified as a magnetic phase distinct from both ferromagnetism and antiferromagnetism (AF)~\cite{Smejkal2022Sep, Smejkal2022Dec, Mazin2022Dec, Mazin2022Dec, Jungwirth2025, Song2025}. Altermagnetic order breaks time-reversal symmetry \tr by itself, but preserves the combination of \tr with some crystallographic symmetries, ensuring compensation of the net magnetic dipole moment. These symmetries are often non-symmorphic, and cannot be inversion or translation, unlike in AF. This causes otherwise spin-degenerate bands to split in most of the Brillouin zone even in the absence of spin-orbit coupling (SOC), while remaining degenerate on particular nodal lines and planes. Analogous to the nodal structure found in unconventional singlet superconductors, these spin-splitting patterns in altermagnets are labeled as d-wave, g-wave, etc., corresponding to angular momentum $l=2$, $l=4$, etc, which in turn determine the number of nodal planes in the spin-splitting throughout the Brillouin zone~\cite{Antonenko2024Feb, Jungwirth2024Sep, Roig2024Oct, Fernandes2024Topological}.

This connection between nodal behavior and angular momentum motivated the identification of multipolar order parameters for altermagnetism~\cite{Suzuki2017Mar, Hayami2019Nov, McClarty2024Apr, Leeb2024Jun, Radaelli2024Dec}. On the experimental side, there have been efforts and proposals to probe magnetic multipoles through perturbations that assess matrix elements in observable tensors like piezomagnetism \cite{Aoyama2024Apr, Yershov2024May}, second-order magnetoelectric effect~\cite{Zyuzin2024Jun, Smejkal2024Nov, Oike2024Nov, Sun2024Dec} and the magneto-optic Kerr effect~\cite{Gray2024Apr, Iguchi2024Sep}. 
On the theoretical side, magnetic multipoles have been studied in metals via effective momentum-space models in connection with Pomeranchuk instabilities \cite{Pomeranchuk1958,Oganesyan2001,Wu2007}, and in insulators via an atomic description of $f$ or $d$  electron levels subjected to crystalline fields and SOC \cite{Santini2009,Paramekanti2020,Voleti2020}.
However, visualization and calculation of AM multipolar order parameters from first principles in real materials in real space is rather rare~\cite{Bhowal2024Feb, Verbeek2024May, Jaeschke-Ubiergo2025Mar}, and their behavior under SOC has not been studied in detail. 

In this Letter, we approach this problem using a combination of group theory and density functional theory (DFT) calculations. First, we explicitly demonstrate that typical collinear altermagnets such as MnF$_2$ and CrSb host higher-order atomic multipoles in real space, including not only octupoles but dotriacontapoles (32-poles), which are visible in the spatially resolved noncollinear spin density (NSD). 
Inspired by recent proposals that the microscopic origin of AM in MnF$_2$ is the ferro-octupolar order~\cite{Bhowal2024Feb}, and in $\alpha$-Fe$_2$O$_3$ the ferro-32-polar order~\cite{Verbeek2024May}, we present a method to discern atomic magnetic multipoles allowed by symmetry, predict their relative order and quantify their magnitude. While similar methods~\cite{Merkel2023} have been used to, e.g. quantify atomic magnetoelectric multipoles in actinides~\cite{Bultmark2009Jul} and monopoles in lithium transition metal phosphates~\cite{Spaldin2013Sep}, we visualize the spin density and consider the effect of SOC in collinear AMs. In particular, we elucidate the connection of SOC to the NSD, and the induced multipole components. 
The NSD, while smaller than the collinear component by more than an order of magnitude, exhibits a nodal structure with distinct higher-order multipolar behavior consistent with symmetry, providing the most direct illustration of the AM order. We also argue that these higher-order multipoles are relevant to the macroscopic response. We present results for prototypical AMs MnF$_2$ and CrSb, as well as  perovskite KMnF$_3$, where structural degrees of freedom --the fluorine octahedral rotations-- tune the magnetic state from simple AF to AM, thus highlighting the interplay of magnetism with crystallographic symmetries in AMs.

\section{Collinear magnetic multipoles}
Magnetically ordered phases of crystals can be classified using magnetic space groups (MSGs), which also determine the magnetic site symmetries, i.e. the magnetic point group of each atomic site. Similar to the Wyckoff positions commonly used in crystallography, each atomic site can be assigned a magnetic Wyckoff position, which are tabulated for all MSGs \cite{Perez-mato2015, Gallego2012}. 
While the magnetic Wyckoff position tables focus on the allowed dipole moments, the magnetic site symmetries also constrain the spin density $\textbf{m}(\textbf{r})$ around an atomic site. 
Generally, $\textbf{m}(\textbf{r})$ can be expanded in tesseral (cubic) harmonics of order $\ell$, forming a magnetic multipole of order $\ell+1$~\cite{Jackson1998Aug, Hayami2018Oct}. Hence, the allowed magnetic multipoles centered on an atomic site are determined by the site symmetry as well.  
However, the sign and magnitude of multipoles are not determined by the site symmetry. Below, we obtain the spin density $\textbf{m}(\textbf{r})$ in real space from DFT, and calculate the magnitude of multipoles by projecting its components within atomic spheres onto different tesseral harmonics \cite{Supplement} (see, e.g. Ref. \cite{Mosca2022, Bhowal2024Feb, Jaeschke-Ubiergo2025Mar} for related approaches). These magnitudes are given in the atomic units of $a_0^\ell \mu_B$, where $a_0$ is the Bohr radius and $\mu_B$ the Bohr magneton.
In the following, we use a shorthand notation and, for example, denote the octupole with a spin density pattern along $z$ that depends on position $\textbf{r}$ as $m_z(\textbf{r})\sim xy$ as  `$d_{xy} m_z$'. The magnitude of this octupole is $\sim\int xy\cdot m_z(\textbf{r})~d\textbf{r}$, where the integral is over the atomic sphere (see SM for details~\cite{Supplement}).
This approach does not take into account all contributions to the total unit cell multipoles, but it allows predicting which atomic multipoles are in-phase (IP) vs. out-of-phase (OP) on different atoms within a unit cell. Thus, it circumvents the issue that unit cell multipoles do not have well-established gauge-invariant formulations due to complications in defining the position operator in periodic systems \cite{Resta1994, Spaldin2013Sep}.

\begin{figure}
    \centering
    \includegraphics[width=0.99\linewidth]{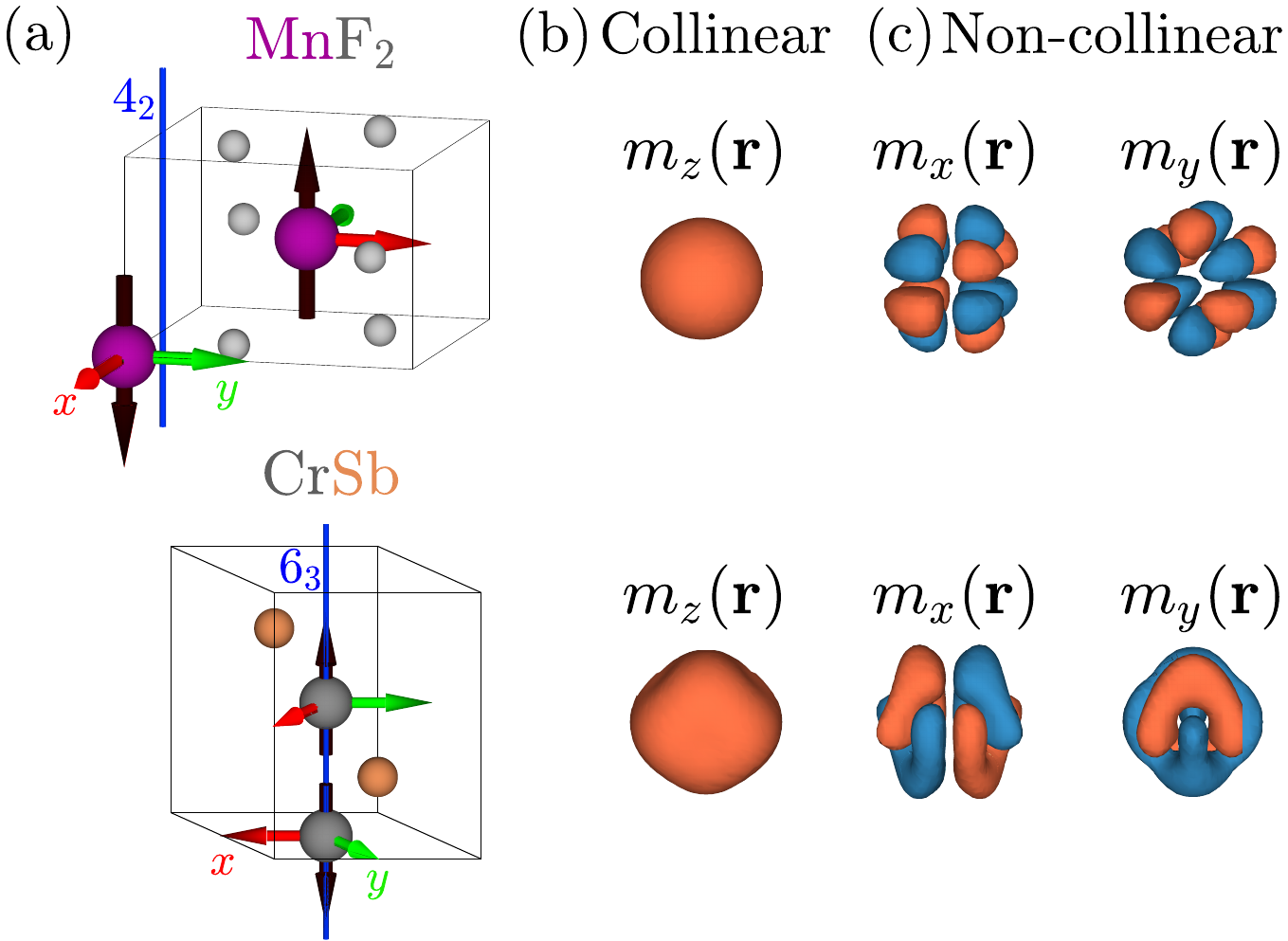}
    \caption{(a) Crystal and magnetic structures of MnF$_2$ and CrSb. The blue lines represent the screw axes, the red and green arrows represents the local $x$, $y$ coordinate axes, and the black arrows represent the spins. (b) Isosurfaces of collinear spin density $m_z(\mathbf{r})$ around a magnetic ion obtained from DFT in the absence of SOC. When the SOC is introduced, the atomic dipoles do not tilt, but a nodal spin density is induced in the $m_x(\mathbf{r})$, $m_y(\mathbf{r})$ components, displayed in (c), with a smaller isosurface value than (b).}
    \label{fig:structures}
\end{figure}

In all systems we consider, the atomic dipole moments $\int\textbf{m}(\textbf{r})d\textbf{r}$ are along the crystallographic $c$ axis, and the spin density along this direction ($m_z(\textbf{r})$) is orders of magnitude larger than other components. Thus, if one focuses only on the dipole moments at the magnetic sites, the system is collinear. We thus refer to magnetic multipoles induced by $m_z(\textbf{r})$ as \textit{collinear} multipoles, and focus on them first. Rutile MnF$_2$ has the same magnetic structure as was originally proposed for RuO$_2$, with anti-parallel Mn spins oriented along the $c$ axis as shown in Fig.~\ref{fig:structures}a, giving rise to an effective ferro-octupolar order \cite{Bhowal2024Feb}. The MSG is $P4_2'/mnm'$ (\#136.499), and the order parameter transforms as the $m\Gamma_2^+$ space group irrep of the paramagnetic group $P4_2/mnm.1'$ (\#136.496). Mn ions occupy the magnetic Wyckoff site 2a with site symmetry $mm'm'$, which allows  $s m_z=\int m_z(\mathbf{r})d\mathbf{r}$ atomic dipoles, as well as higher order magnetic multipoles such as the $d_{xy} m_z$ octupole, even though the spin density $m_z(\textbf{r})$  from DFT is almost spherical (Fig.~\ref{fig:structures}b).

Site symmetry alone does not distinguish between antiferromagnets and altermagnets: if all atomic octupoles had alternating signs on neighboring Mn ions, they would have given rise to antiferromagnetism only. For MnF$_2$ to be altermagnetic, at least one type of octupole on the two Mn ions in the unit cell have to be in-phase (in a global coordinate system) to add up to a macroscopic octupole moment \cite{Bhowal2024Feb}.

In the Supplemental Material (SM) \cite{Supplement}, we provide details of the irrep subduction analysis \cite{Elcoro2017Oct}, which shows that the only multipoles allowed on Mn ions, apart from the $A_g^-$ ones, transform as the $B_{1g}^-$ irrep of the $mmm.1'$ point group; i.e. $m\Gamma_2^+\downarrow mmm.1'\rightarrow A_{g}^- \oplus B_{1g}^-$. As shown in Fig.~\ref{fig:structures}a, in MnF$_2$ there is a nonsymmorphic $4_2$ screw symmetry that interchanges the two Mn atoms in the unit cell, and as a result of this symmetry, the local coordinate axes of these Mn ions are rotated by 90$^\circ$ degrees with respect to each other. This means that the local multipoles on either Mn ion are rotated with respect to each other, and even though all $B_{1g}^-$ multipoles on the two Mn ions are out-of-phase in the local coordinate systems, they are in-phase in a \textit{global} coordinate system if they change sign under 90$^\circ$ rotation around the $c$ axis ($C_{4z}$). 
This in turn is determined from the angular dependence of the multipoles; e.g. $d_{xy}m_z$ is odd under $C_{4z}$ because $m_z$ is even under it but the orbital part $d_{xy}$ changes sign.
The allowed collinear multipoles grouped according to whether they are in-phase (IP) or out-of-phase (OP) in \textit{global} coordinates are: 
\begin{align*}
    (\rm{OP})\, m\Gamma_2^+\, [B^-_{1g}] &: s m_z, d_{z^2}m_z,\; g_{z^4}m_z,\; g_{x^2y^2}m_z, \\
    (\rm{IP})\, m\Gamma_2^+\, [B^-_{1g}] &: d_{xy}m_z,\; g_{xyz^2}m_z. 
\end{align*}
Here, $gm_z$ are 32-poles with $m_z(\textbf{r})$ forming a $g$-orbital-like ($l=4$) nodal pattern. 
The collinear dipoles are OP and hence cancel each other, as is expected for a compensated magnet, together with all the other OP multipoles. On the other hand, the octupole $d_{xy}m_z$ and the 32-pole $g_{xyz^2}m_z$ are IP and do not cancel. 
This is in line with the observation that collinear compensating moments in the rutile structure yield a d-wave altermagnet, i.e. the lowest non-zero multipole in the unit cell is an octupole~\cite{Bhowal2024Feb}.

Unlike MnF$_2$, the hexagonal metal CrSb with MSG $P6_3'/m'm'c$ is a g-wave altermagnet \cite{Reimers2024Mar, Li2024May, Ding2024Nov}. Even though the site symmetry $\bar{3}m'$ allows collinear octupoles on the Cr ions in its AM phase, all of them are out-of-phase in the global coordinate system, and the lowest order in-phase multipole is the $g_{yz(y^2-3x^2)}m_z$ dotriacontapole: 
\begin{align*}
    (\rm{OP})\,m\Gamma_4^+\, [A^-_{2g}] &: s m_z,\; d_{z^2}m_z,\; g_{z^4}m_z, \\
    (\rm{IP})\,m\Gamma_4^+\, [A^-_{2g}] &: g_{yz(y^2-3x^2)}m_z,
\end{align*}
In other words, both the collinear dipoles and the octupoles cancel, and the lowest order collinear magnetic multipole that remains uncompensated is a 32-pole. Again, this is expected since the two magnetic ions are swapped by either the $6_3$ screw, shown in Fig.~\ref{fig:structures}b, or a $c$-glide. The only multipole that is IP is the one whose orbital part is odd under both the 6-fold rotation and the diagonal $\{110\}$ mirrors.

\begin{figure}
    \centering
    \includegraphics[width=0.95\linewidth]{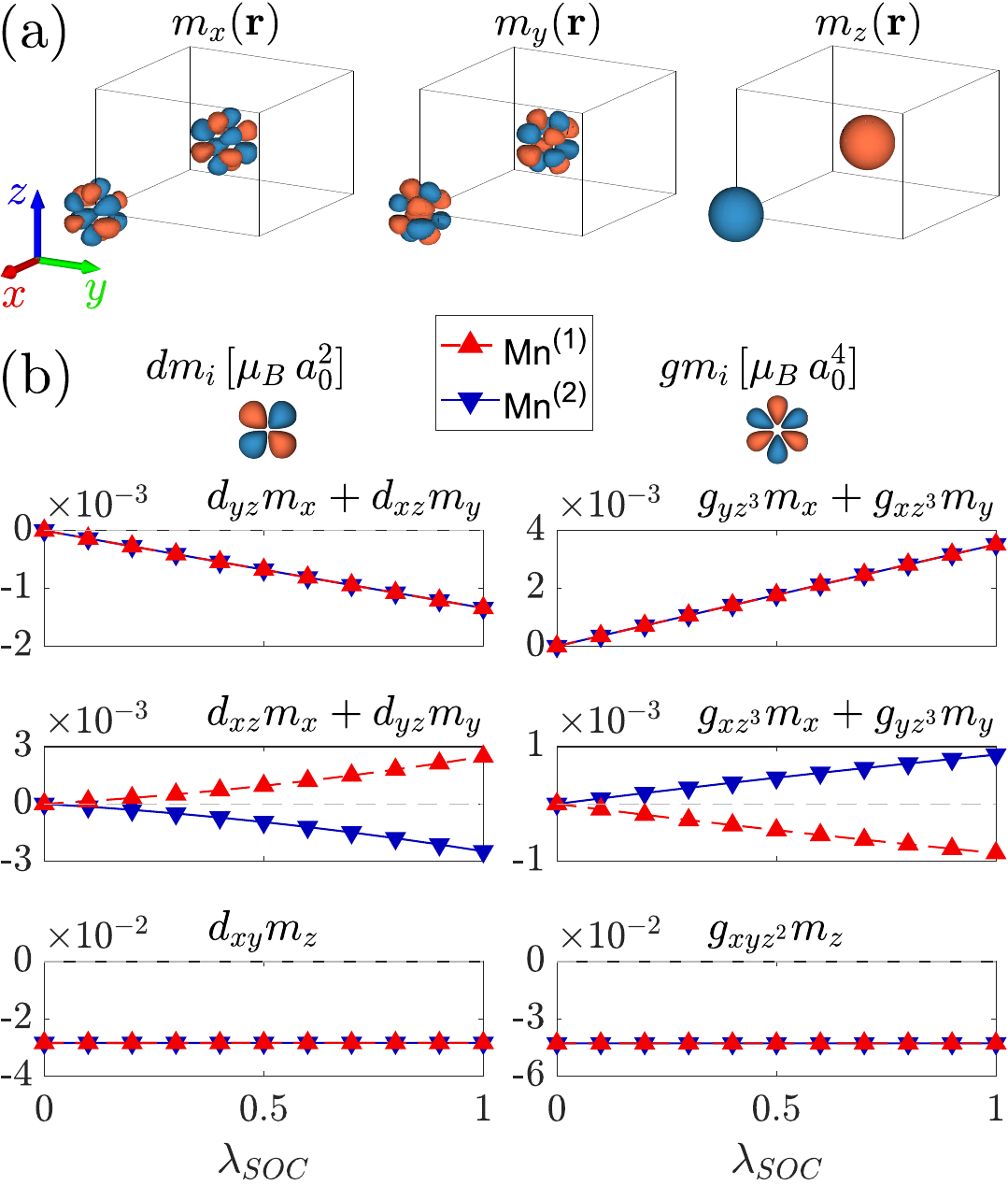}
    \caption{(a) The isosurfaces for different components of DFT spin densities of MnF$_2$. Orange and blue colors correspond to positive and negative values, and the isosurface values used for collinear ($m_z$) and noncollinear ($m_x$ and $m_y$) are not equal. (b) Examples of decompositon onto the tesseral harmonics of the spin density in MnF$_2$. The atomic magnetic octupoles have the units of $\mu_Ba_0^2$, while the 32-poles have the units of $\mu_Ba_0^4$, where $\mu_B$ is the Bohr magneton and $a_0$ is the Bohr radius. Red and blue data correspond to the Mn atoms on the corners and the body-center of the unit cell, which are equal for IP multipoles (e.g. $d_{xy}m_z$) and opposite for OP multipoles (e.g. $m_z$ and $d_{xz}m_x+d_{yz}m_y$). Results for the rest of the multipoles are reported in the SM \cite{Supplement}.}
    \label{fig:MnF2}
\end{figure}

\section{Spin-orbit coupling} 
\begin{figure}
    \centering
    \includegraphics[width=0.99\linewidth]{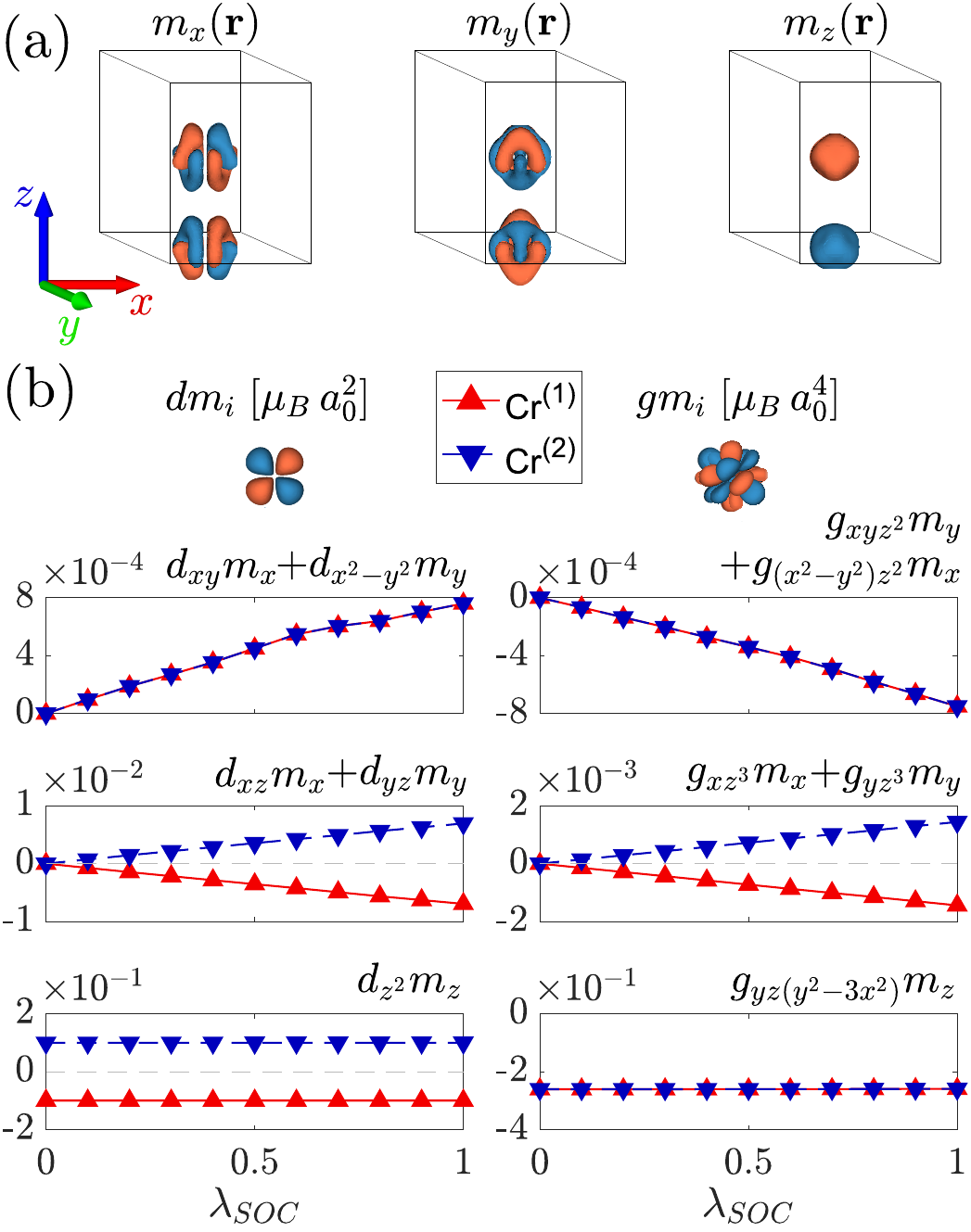}
    \caption{(a) The isosurfaces for different components of DFT spin densities of CrSb. Orange and blue colors correspond to positive and negative values, and the isosurface values used for collinear ($m_z$) and noncollinear ($m_x$ and $m_y$) are not equal. (b) Examples of decompositon onto the tesseral harmonics of the spin density in CrSb. The atomic magnetic octupoles have the units of $\mu_Ba_0^2$, while the 32-poles have the units of $\mu_Ba_0^4$, where $\mu_B$ is the Bohr magneton and $a_0$ is the Bohr radius. Red and blue data correspond to the Cr atoms on the edge and the body-center of the unit cell, which are equal for IP multipoles (e.g. $d_{xy}m_x+d_{x^2-y^2}m_y$) and opposite for OP multipoles (e.g. $d_{z^2}m_z$). Results for the rest of the multipoles are reported in the SM \cite{Supplement}.}
    \label{fig:CrSb}
\end{figure}

Both MnF$_2$ and CrSb are collinear magnets with atomic dipoles parallel to the $c$ axis. However, any magnetic material can have a noncollinear spin density (NSD) in the general equivalent positions $(x,~y,~z)$ in the unit cell, because these positions do not have any symmetries that prohibit a local spin density $\textbf{m}(\textbf{r})$ with an arbitrary orientation. This fact, often ignored because the NSD cancels out when integrated around an atomic site in collinear magnets ($sm_x=\int m_x(\textbf{r}) d\textbf{r} =sm_y=0$), is easy to demonstrate in DFT by simply plotting isosurfaces of the NSD. In Fig.~\ref{fig:structures}c, we show the isosurfaces of $m_x(\textbf{r})$ and $m_y(\textbf{r})$ spin densities around Mn and Cr ions in MnF$_2$ and CrSb from DFT. Even though MnF$_2$ is a d-wave AM, it exhibits 32-poles in its real space NSD that do not cancel when summed over the two atoms, as shown in ~\ref{fig:MnF2}a. This means that even though the primary order paramater is octupolar, higher-order in-plane multipoles are dominant and affect physical observables that depend on in-plane moments. In CrSb, while non-canceling high-order multipolar behavior is also evident (see \ref{fig:CrSb}a), the spin density pattern is more complicated because of the partial occupation of the Cr orbitals. This necessitates projecting the NSD onto tesseral harmonics to single out the contributions from different multipoles.

We now consider the effect of SOC on the NSD. While the NSD is allowed by MSG symmetries, it is not invariant under a spin-only rotation $R_s(\theta)$ by and arbitrary angle $\theta$ around the  $c$ axis. As a result, when SOC is ignored and $R_s(\theta)$ becomes a symmetry of the system, the noncollinear multipoles become zero. 
In Figures~\ref{fig:MnF2}b and \ref{fig:CrSb}b, we present the magnitude of magnetic multipoles in MnF$_2$ and CrSb, calculated from first principles, as a function of the SOC strength. In these figures, $\lambda_{SOC}=0$ corresponds to SOC being turned off, and $\lambda_{SOC}=1$ corresponds to the real value of SOC. 
When there is no SOC, the NSDs $m_x$, $m_y$ go to zero. Under finite SOC, the collinear multipoles are modified quadratically in $\lambda_{SOC}$, and only by a small amount that is unnoticeable in the scale of the figures~\cite{Supplement}. On the other hand, while the noncollinear dipoles always remain zero by symmetry, the symmetry-allowed noncollinear octupoles and 32-poles become non-zero linearly in $\lambda_{SOC}$. Whether these components on the two different ions in the unit cell are in-phase or out-of-phase is consistent with the group theory predictions discussed earlier.

In MnF$_2$, the noncollinear multipoles on the Mn ions allowed by symmetry are
\begin{align*}
    (\rm{OP})\, m\Gamma_2^+\, [B^-_{1g}] :\; &d_{xz}m_x+d_{yz}m_y, \\
    &g_{xz(x^2-3y^2)}m_x+g_{yz(3x^2-y^2)}m_y, \\
    (\rm{IP})\, m\Gamma_2^+\, [B^-_{1g}] :\; &d_{yz}m_x+d_{xz}m_y, \\
    &g_{yz(y^2-3x^2)}m_x+g_{xz(x^2-3y^2)}m_y.
\end{align*}
As was the case for the collinear ones, the noncollinear multipoles that are IP are the ones that are odd under 4-fold rotation, which is confirmed by the DFT results. 
In the $g$-wave altermagnet CrSb, non-zero multipoles on the Cr ions are
\begin{align*}
    (\rm{OP})\,m\Gamma_4^+\,[A^-_{2g}] :\; &d_{yz}m_x + d_{xz}m_y,\; g_{yz^3}m_x + g_{xz^3}m_y, \\
    (\rm{IP})\,m\Gamma_4^+\,[A^-_{2g}] :\; &d_{xy}m_x + d_{x^2-y^2}m_y, \\
                 &g_{xyz^2}m_x + g_{(x^2-y^2)z^2}m_y, \\
                 &g_{xy(x^2-y^2)}m_x + g_{x^2y^2}m_y,
\end{align*}
where, surprisingly, the lowest order IP noncollinear multipoles that do not cancel are octupoles that are odd under the 60$^\circ$ rotation. 

The linear dependence of the noncollinear multipoles on $\lambda_{SOC}$ can be explained by the form of the commutator of $\mathcal{H}_{SOC} \sim \lambda_{SOC} \textbf{L}\cdot \textbf{S}$ with the unperturbed Hamiltonian $\mathcal{H}_0$, in terms of the orbital and spin ladder operators $L_\mp$ and $S_\mp$. This commutator is off-diagonal both in spin and orbital spaces, so starting from a state $|\Psi_0\rangle$ that is an eigenstate of $S_z$, the first order change in the wavefunction only includes terms with spin opposite to $|\Psi_0\rangle$. Hence, there is an in-plane spin density at first order in $\lambda_{SOC}$. However, the change in the non-spin-flip part of the wavefunction is nonzero only in the second order in the perturbation, which explains the trend observed in the collinear multipoles above. The emergence of a spin density with a $g$-orbital-like pattern in a system with spin-polarized electrons on the $d$-shell can also be explained similarly. In the SM \cite{Supplement}, we provide a simple toy model to illustrate these observations more explicitly.

\section{Structural control of altermagnetism }
Finally, we discuss a third material example to elucidate how IP multipoles emerge in an antiferromagnetic-to-altermagnetic transition. The perovskite crystal structure \cite{Fernandes2024Topological,Naka2025,Bernardini2025} is an extremely flexible platform to realize antiferromagnetism and altermagnetism. In particular, the perovskite KMnF$_3$ \cite{Naka2025} provides a suitable playground to study the dependence of altermagnetism on structural degrees of freedom. This compound undergoes a structural transition at $\sim$88~K from cubic $Pm\bar{3}m$ to tetragonal $I4/mcm$ space group. G-type collinear magnetic order also condenses at either the same or at a close temperature \cite{Carpenter2012Jun, Knight2020Nov}. The magnetic order parameter transforms as the $mR_5^-$ irrep of the parent $Pm\bar{3}m.1'$ space group. Being a zone-boundary order, it preserves anti-translation (translation followed by time reversal) symmetry and hence is an AF that does not give rise to no spin-splitting.

\begin{figure}
    \centering
    \includegraphics[width=0.99\linewidth]{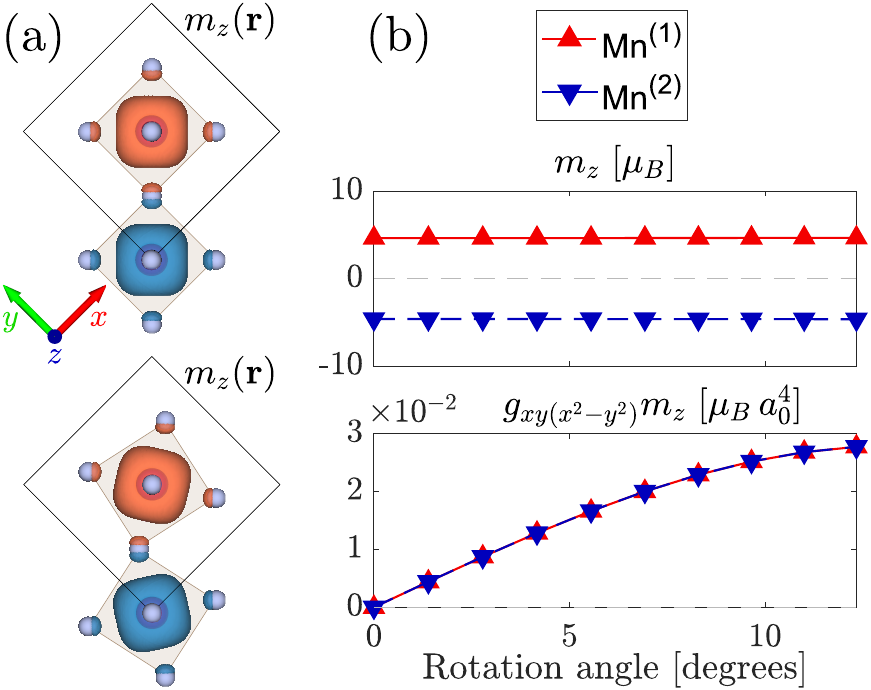}
    \caption{(a) The $m_z(\mathbf{r})$ spin density around the Mn ions in KMnF$_3$ is aligned with the F octahedra, and rotates without much distortion along with the octahedra. F ions are shown as grey spheres. (b) 
    While the integrated $sm_z$ atomic dipole density remains almost constant as a function of the octahedral rotation angle, the IP 32-poles $g_{xy(x^2-y^2)}m_z$ turn on linearly with octahedral rotations. All other multipole components are reported in the SM \cite{Supplement}.}
    \label{fig:KMnF3}
\end{figure}

In the tetragonal phase, the fluorine octahedra are rotated by $\sim 6 ^o$ around the $c$ axis, and the unit cell is doubled. The oxygen octahedral rotation order parameter transforms as the $R_5^-$ irrep, and has the same wavevector as the magnetic order parameter. In other words, it breaks the translational symmetry the same way that the magnetic order parameter does, and hence anti-translation is no longer a symmetry when both structural and magnetic order parameters condense. This leads to \textit{octahedral rotation induced altermagnetism} in KMnF$_3$, where the altermagnetic splitting is proportional to both the magnetic order parameter and the structural order parameter. 
This is evident in the DFT results. In Fig.~\ref{fig:KMnF3}, we show that the $m_z(\textbf{r})$ isosurfaces around each Mn ion rotate with the octahedra, and even though the total dipole density does not change much, the lowest order IP multipole, the 32-pole $g_{xy(x^2-y^2)}m_z$, is linearly proportional to the octahedral rotation angle (Fig.~\ref{fig:KMnF3}b). 
This 32-pole transforms as the $m\Gamma_1^+$ irrep of $Pm\bar{3}m.1'$, and it is induced by a trilinear coupling in the free energy: $\mathcal{F}\sim R_5^-(a,0,0)\cdot mR_5^-(a,0,0) \cdot m\Gamma_4^+$, consistent with its linear dependence on $R_5^-$ rotations \cite{Supplement}. A similar rotation-induced altermagnetic phase with the same rotation mode (albeit with a rhombohedral structure) has also been studied in BiFeO$_3$~\cite{Urru2025}, attesting to the fact that it is a common feature in many magnetic perovskites. 

We note that the related compound RbMnF$_3$ has a N\'eel temperature that is comparable to that of KMnF$_3$, but it retains the cubic crystal structure at all temperatures \cite{Ortiz2014}. As a result, the partially Rb-substituted compound K$_{1-x}$Rb$_x$MnF$_3$ would likely have a lower temperature structural transition that is well separated from the magnetic one, and hence could serve as a model system to study antiferromagnetic to altermagnetic phase transitions. 

\section{Tensor observables}
The symmetry of response tensors is often described using the Jahn symbol \cite{Jahn1949Mar}. The multipoles can also be connected to a Jahn symbol~\cite{Radaelli2024Dec, Schiff2025}: for example, the magnetic octupole $r_ir_jm_k$ is an axial and magnetic rank-3 object symmetric under exchange of $i$ and $j$, so it is denoted by the symbol $ae[V^2]V$. Thus, systems with a net magnetic octupole must allow non-zero elements of  $ae[V^2]V$ response tensors. Examples of such tensors include piezomagnetism~\cite{Dzialoshinskii1958Jan, Moriya1959Sep} and the second-order magnetoelectric effect \cite{Gallego2019May, Urru2022Dec}. 

\begin{figure}
    \centering
    \includegraphics[width=0.99\linewidth]{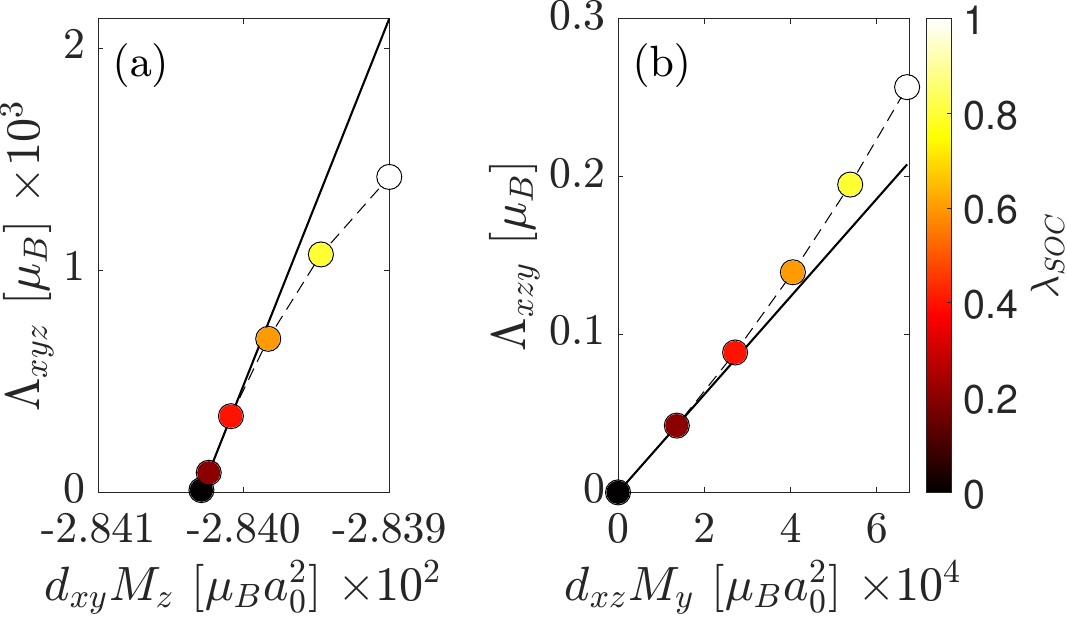}
    \caption{The dependence of the piezomagnetic tensor components to the corresponding magnetic octupoles in MnF$_2$. Different data points are generated by repeating the DFT calculations for different SOC strengths. (a) $\Lambda_{xyz}$, which relates the $z$ component of magnetization to $xy$ shear strain, is symmetry allowed even in the absence of SOC, however, it is forced to be zero by Luttinger compensation at zero temperature since there is a gap. However, the corresponding atomic octupole $d_{xy}M_z$ is not affected by the compensation and is nonzero for all values of SOC. The interplay between the compensation and response leads to a nontrivial dependence of this component of the piezomagnetic tensor on the atomic multipole strength. (b) For small values of SOC, the $\Lambda_{xzy}$ component, which relates the $y$ component of magnetization to $xz$ shear strain, linearly depends on the $d_{xz}M_y$ magnetic octupole, and both quantities are zero in the absence of SOC.}
    \label{fig:piezo}
\end{figure}

The IP multipoles in MnF$_2$ include the collinear component $d_{xy}m_z$, which corresponds to the piezomagnetic tensor element $\Lambda_{xyz}$, which we define through the linear relationship between the applied strain $\epsilon_{ij}$ and the induced magnetic moment $M_k$ as 
\begin{equation}
M_k = \Lambda_{ijk} \epsilon_{ij}. 
\end{equation}
Since $\Lambda_{xyz}$ is induced by the large collinear spin components, this element remains finite even in the absence of SOC, but it is strongly suppressed in this system by Luttinger compensation, which enforces equal numbers of spin up and down electrons at zero temperature as long as the gap remains open~\cite{Mazin2022Dec, LinDing2024, Khodas2025}. 
In order to make this prediction more evident, in Fig.~\ref{fig:piezo}(a) we plot the magnitude of $\Lambda_{xyz}$ as a function of the multipole $d_{xy}m_z$. We calculated the piezomagnetic coefficients by  straining the unit cell in our DFT calculations, relaxing the atomic positions under fixed strain, and performing a self-consistent electronic structure calculation to obtain the value of magnetization. In the calculation without SOC, the $d_{xy}m_z$ multipole is nonzero but the $\Lambda_{xyz}$ piezomagnetic coefficient is zero within the scale of this plot, as expected. This, however, is not the case for other components of $\Lambda$. For instance, the $\Lambda_{xzy}$ component, which is not symmetry allowed without SOC depends on the corresponding magnetic octupole linearly for small values, with only some deviation from the linear behavior for the actual value of $\lambda_{SOC}$. Note that this component is not affected from the compensation either, because the magnetization it leads to is formed by the tilting of local atomic moments. 

On the other hand, since CrSb is d-wave only in the noncollinear components, strain can induce a linear magnetization only along the $x$ and $y$ directions, and the magnitude of this effect must be proportional to $\lambda_{SOC}$.
The lowest rank tensors that are linearly proportional to the collinear 32-pole $g_{yz(y^2-3x^2)}m_z$ are rank-5 tensors with  $ae[V^2][V^2]V$. These include second-order piezomagnetism (magnetization proportional to the product of two different components of strain, e.g. $\varepsilon_{yz}$ and $\varepsilon_{xx}$), and strain derivative of the second order magnetoelectric effect.

\section{Conclusions}
Using first principles calculations, we calculated the higher order atomic multipoles in altermagnetic materials by projecting the spin density onto multipoles, which provides a simple yet robust approach to calculate these multipoles in real materials and beyond simple models. Interestingly, even in systems with seemingly spherical spin density in the collinear direction, we found that the noncollinear components provide a more elucidative illustration of the presence and relative phase of altermagnetic multipoles. Our results also showed that 32-poles, which are often ignored in the discussion of $d$-wave altermagnets, can be in fact be the dominant contribution to spin density and in certain macroscopic properties. 
In KMnF$_3$, the interplay between the 32-poles and the crystal structure adds altermagnetism to the list of phenomena induced by octahedral rotations in perovskites. 
In summary, our work underlines that the NSD, higher order multipoles, and spin-orbit coupling are all important features of altermagnets that need to be taken into account in real materials especially when predicting their macroscopic response. 

\section{Acknowledgments}
The authors acknowledge useful discussions with Libor \v{S}mejkal and Hana Schiff. L.B. and T.B were supported by the NSF CAREER grant DMR-2046020. L.B. and T.B. were also partially supported 
by the National Science Foundation through the University of Minnesota MRSEC under Award Number DMR-2011401. 
R.M.F. was supported by the Air Force Office of Scientific Research under Award No. FA9550-21-1-0423.

\section{Data Availability} 
The data that support the findings of this article are openly available at the Data Repository for University of Minnesota (Ref.~\cite{drum}).

\end{document}


\title{Supplemental Material - Noncollinear Magnetic Multipoles in Collinear Altermagnets}

\author{Luca Buiarelli}
\affiliation{Department of Chemical Engineering and Materials Science, University of Minnesota, MN 55455, USA}

\author{Rafael M. Fernandes}
\affiliation{Department of Physics, The Grainger College of Engineering, University of Illinois Urbana-Champaign, Urbana, IL 61801, USA}
\affiliation{Anthony J. Leggett Institute for Condensed Matter Theory, The Grainger College of Engineering, University of Illinois Urbana-Champaign, Urbana, 61801, IL, USA}

\author{Turan Birol}
\affiliation{Department of Chemical Engineering and Materials Science, University of Minnesota, MN 55455, USA}

\date{\today}

\maketitle

\section{Methods}
\subsection{DFT}
We simulated all the materials using Density Functional Theory and Projector Augmented Waves (PAW) as implemented in Abinit 9.10.1~\cite{Torrent2008Apr, Marques2012Oct, Gonze2016Aug, Gonze2020Mar, Romero2020Mar}, utilizing JTH v1.1 pseudopotentials~\cite{Jollet2014Apr} within the Local Density Approximation. For MnF$_2$ we used a grid of 20x20x24 k-points centered at $\Gamma$. For KMnF$_3$ we used the conventional body-centered, 20-atom cell and a 12x12x8 k-point grid centered at $\Gamma$. For CrSb we used a grid of 20x20x16 k-points centered at $\Gamma$. For each of the three we used a cutoff energy of 1100 eV for the spherical grid inside the augmentation regions and 550 eV for the normal plane wave basis grid. In order to avoid errors due to structural relaxation in DFT, we used experimentally reported structures for all compounds~\cite{Griffel1950Oct, Takei1963Mar, Knight2020Nov} without any further structural relaxation. The lattice and Wyckoff parameters are reported in Tab.~\ref{tab:structures}. For the insulating MnF$_2$ and KMnF$_3$ we used the insulator occupation scheme of Abinit, while for the metal CrSb we use the metallic occupation scheme with gaussian smearing and an electronic smearing temperature of 0.01 eV.

\begin{table}[]
\begin{tabular}{c|ccc|ccc|c|c}
 & $a$[\AA] & $b$[\AA] & $c$[\AA] & $\alpha$[deg] & $\beta$[deg] & $\gamma$[deg] & F Wycoff paramter & $R$[\AA] \\ \hline
MnF$_2$ & 4.874 & 4.874 & 3.300 & 90 & 90 & 90 & 0.3046 & 1.118 \\
CrSb & 4.103 & 4.103 & 5.463 & 90 & 90 & 120 & - & 1.116 \\
KMnF$_3$ & 5.899 & 5.899 & a$\sqrt{2}$ & 90 & 90 & 90 & 0.2257 & 1.118
\end{tabular}
\caption{Lattice and Wyckoff parameters used in the DFT calculations of the three compounds studied, obtained from experimental characterization~\cite{Griffel1950Oct, Takei1963Mar, Knight2020Nov}. The radius of integration $R$ used for calculating the atomic multipoles is also reported.}
\label{tab:structures}
\end{table}

In MnF$_2$ and CrSb, to study the dependence of the multipoles on spin-orbit coupling, we used the scaling variable provided in Abinit (\textit{spnorbscl}), which we later refer to as $\lambda_{SOC}$. In KMnF$_3$, we study the cubic-to-tetragonal transition by artificially constraining the unit cell to remain cubic (fixing $c = a\sqrt{2}$) while introducing an octahedral rotation. In the conventional cell, this implies changing the Wyckoff parameter of the four Fluorine atoms from $(1/4,3/4,0)$ to $(x,x+1/2,0)$, where $x$ is connected to the octahedral rotation angle $\beta$ by the trigonometric relation
\begin{equation}
    \beta = \cos^{-1}\bigg(\frac{\sqrt{2}x(\frac{1}{2}-x)}{\sqrt{2x^2+\frac{1}{4}-x}}\bigg) .
\end{equation}

\subsection{Magnetic multipoles}
\begin{figure}[H]
    \centering
    \includegraphics[width=0.32\linewidth, clip=true, trim = 250 300 1500 100]{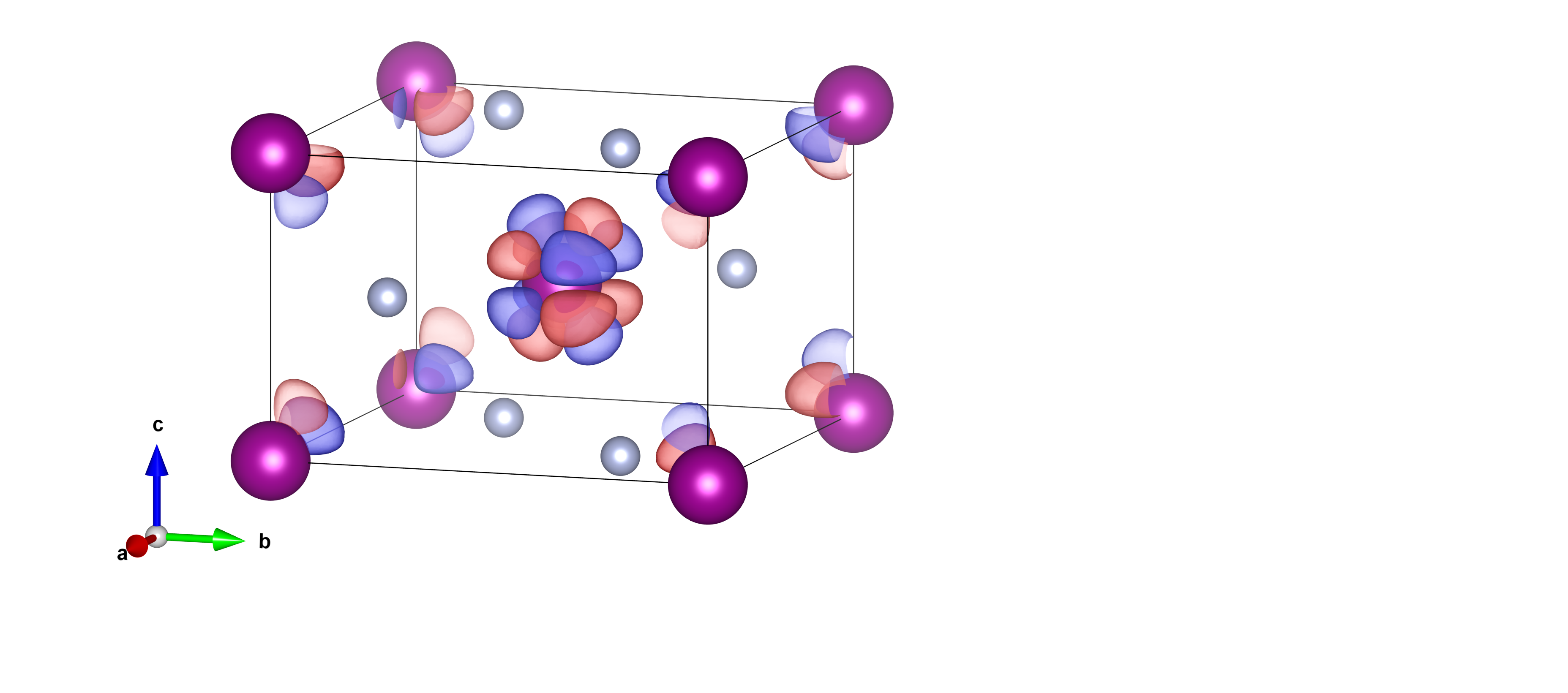} \includegraphics[width=0.32\linewidth, clip=true, trim = 250 300 1500 100]{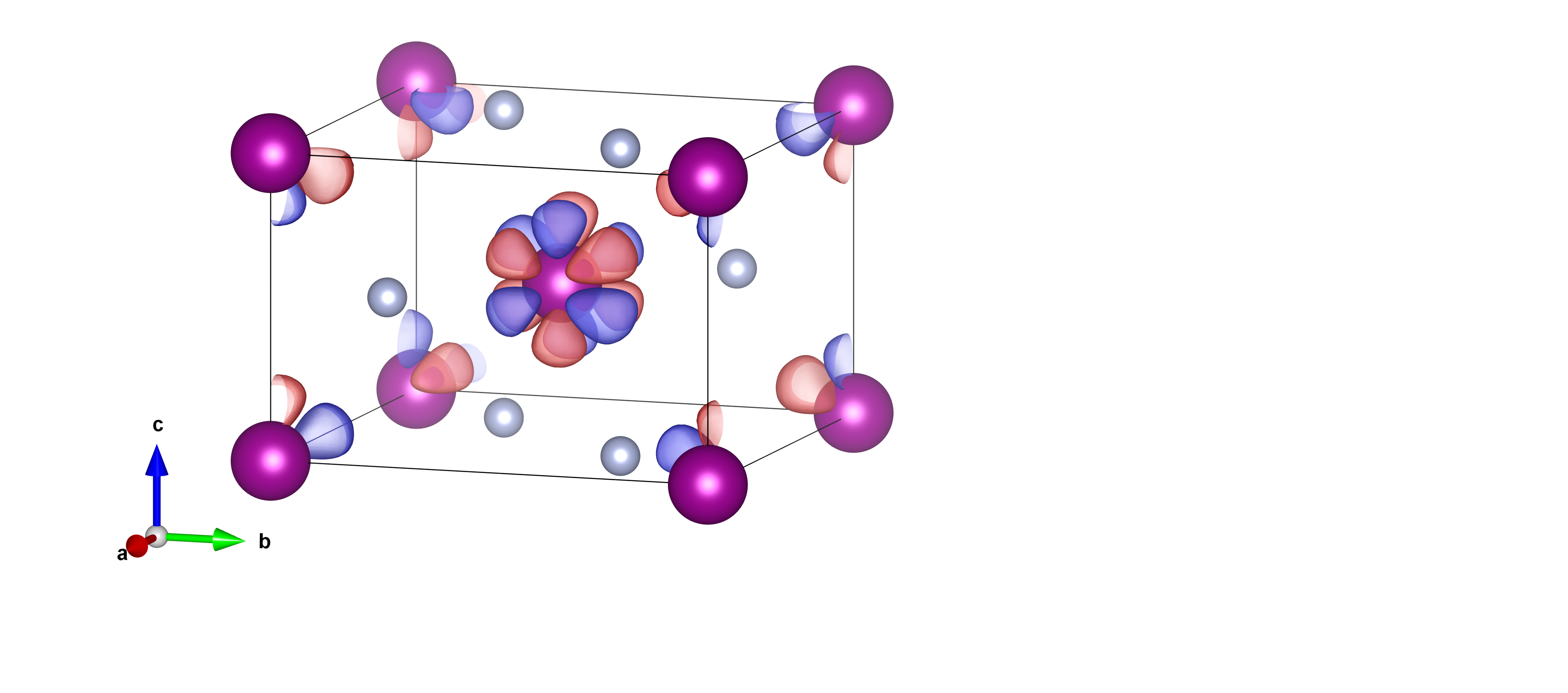} \includegraphics[width=0.32\linewidth, clip=true, trim = 250 300 1500 100]{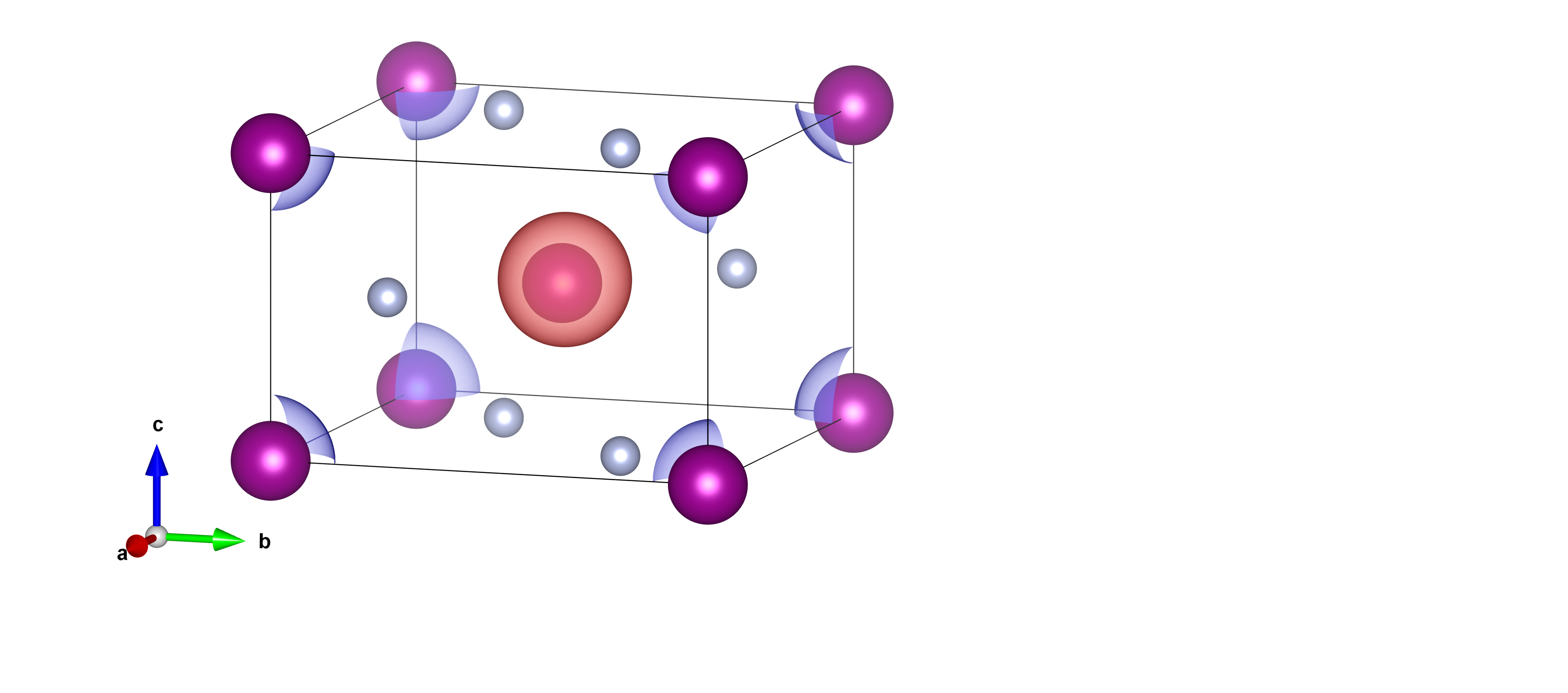}
    \caption{\label{fig:MnF2_spindens} Spin Density of MnF$_2$ projected along the $x$, $y$, and $z$ spin components (left to right). The noncollinear components $m_x$ and $m_y$ show the behavior of a magnetic dotriacontapole (32-pole). Figure generated through the VESTA software~\cite{Momma2008Jun}.}
\end{figure}

\begin{figure}[H]
    \centering
    \includegraphics[width=0.95\linewidth]{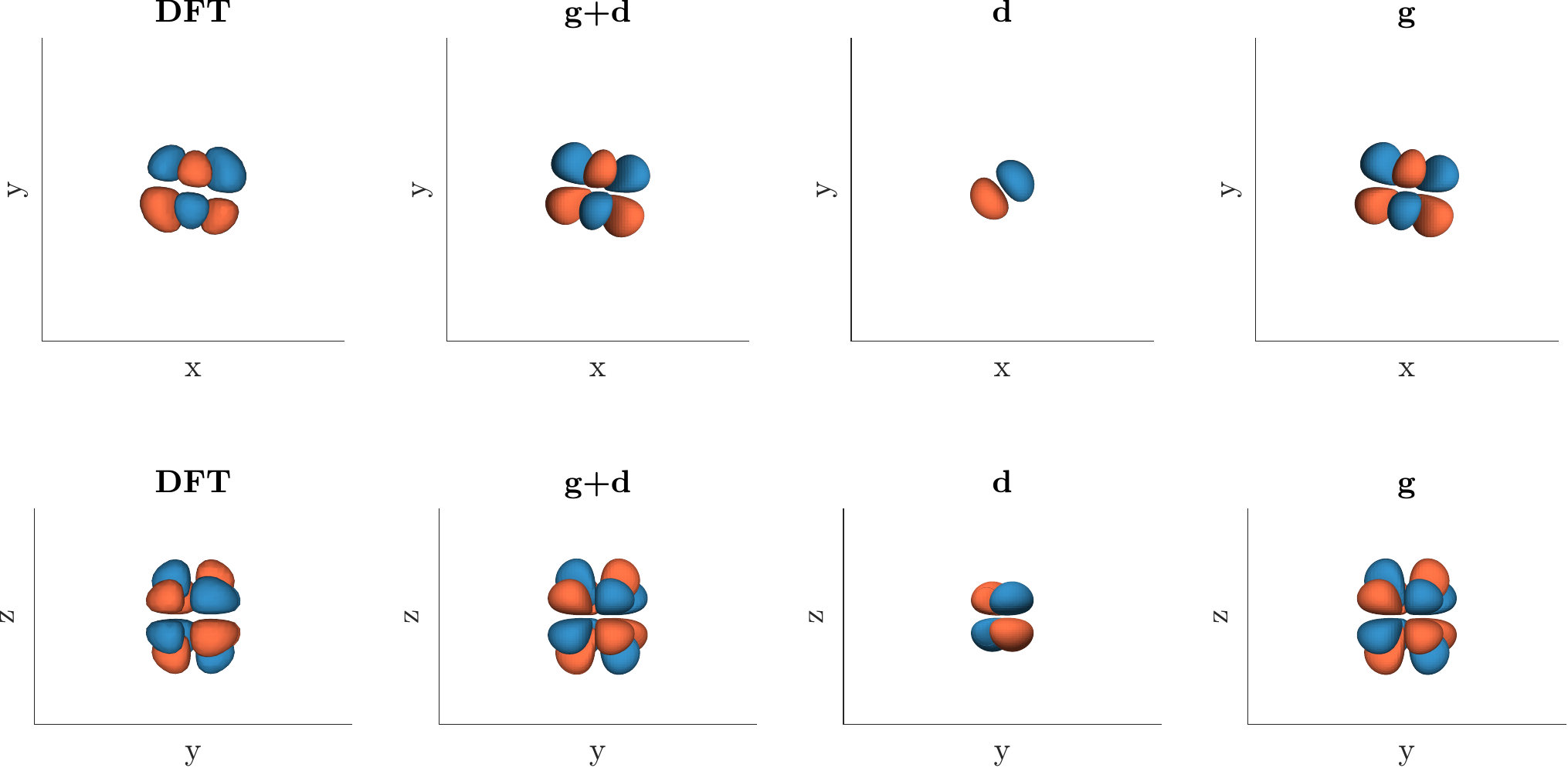}
    \caption{\label{fig:MnF2_mx_SumOfMultipoles} Noncollinear spin density $m_x$ of MnF$_2$ from DFT decomposed onto the tesseral harmonics.}
\end{figure}

\begin{figure}[H]
    \centering
    \includegraphics[width=0.33\linewidth, clip=true, trim = 200 300 1600 50]{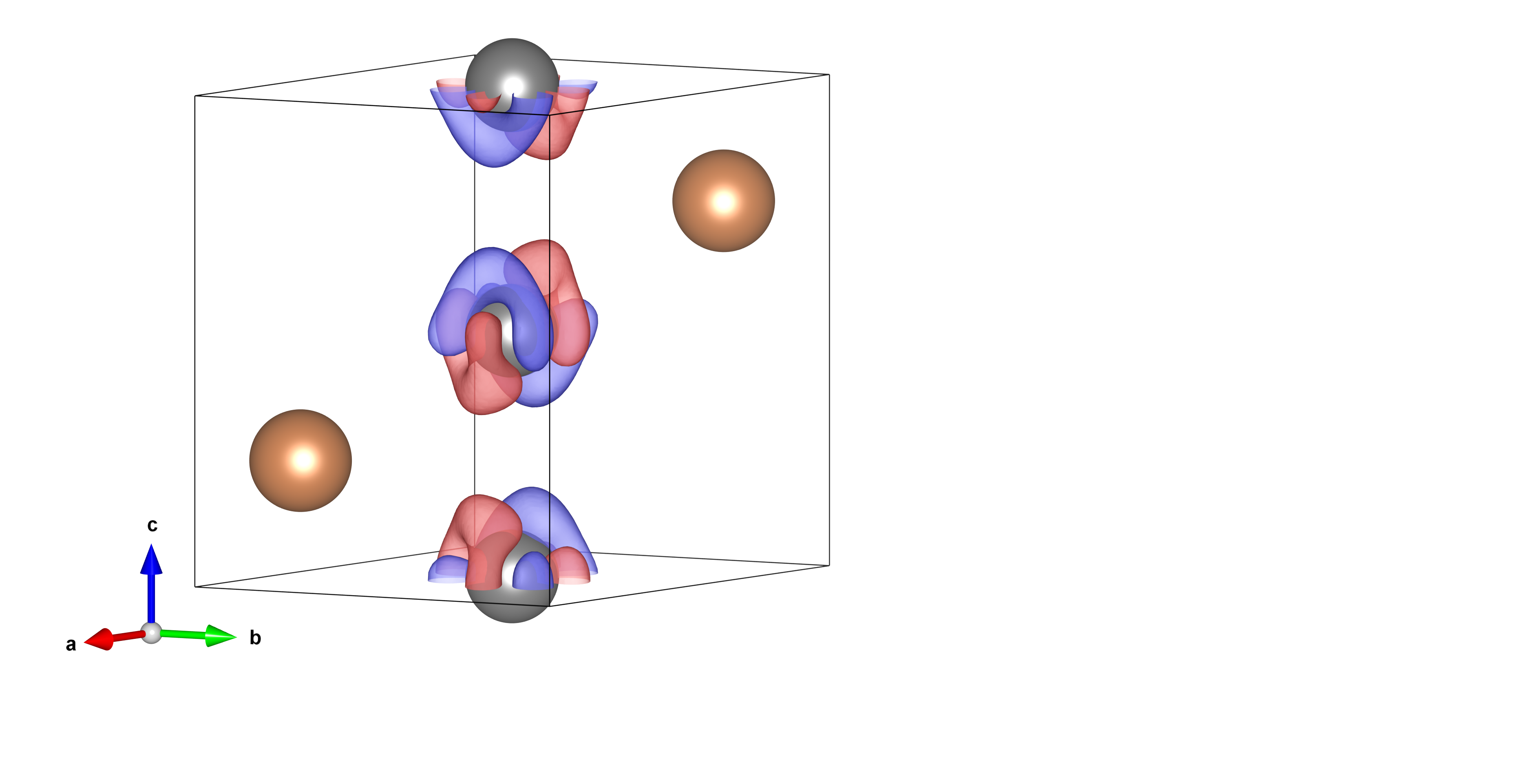}\includegraphics[width=0.33\linewidth, clip=true, trim = 200 300 1600 50]{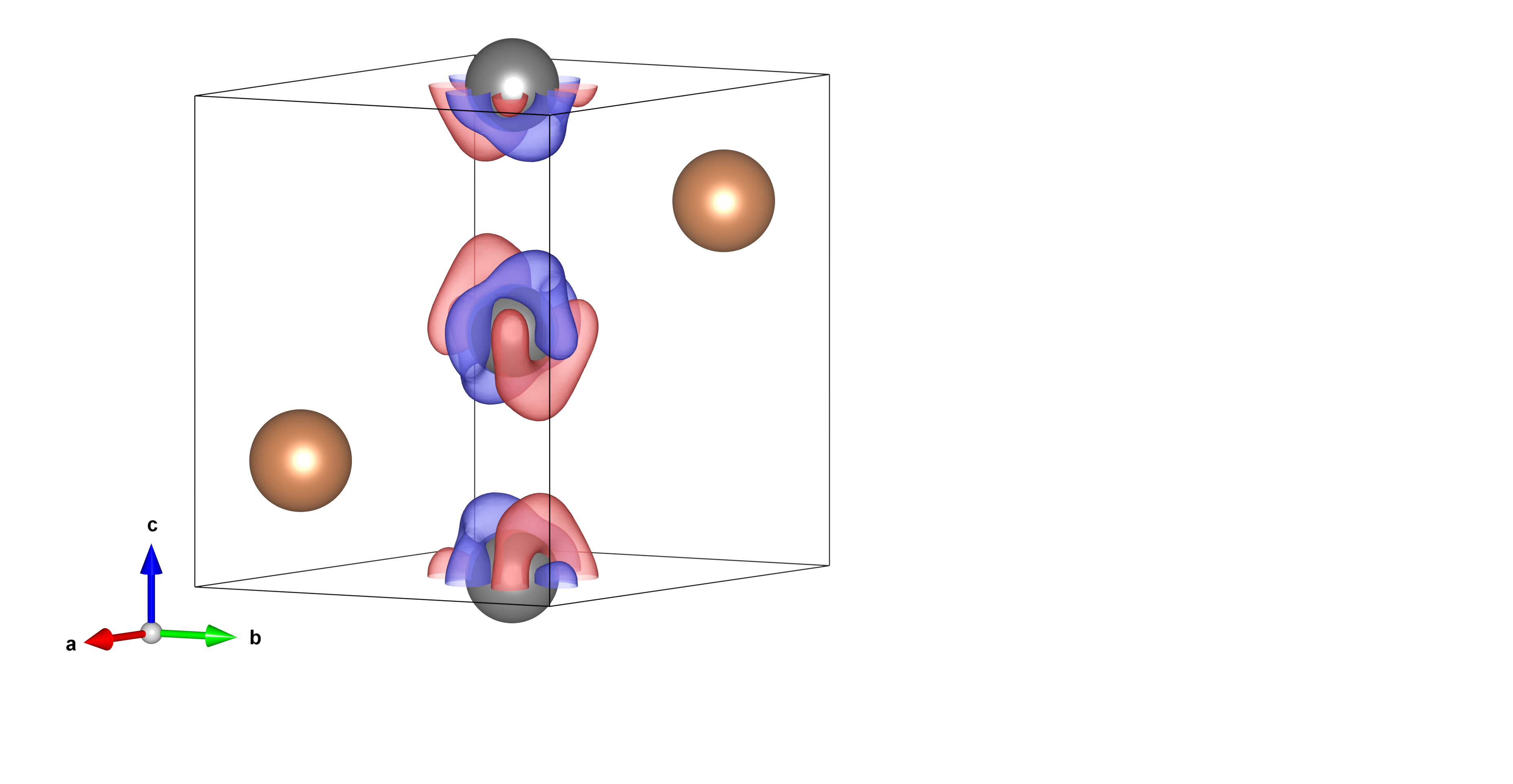}\includegraphics[width=0.33\linewidth, clip=true, trim = 200 300 1600 50]{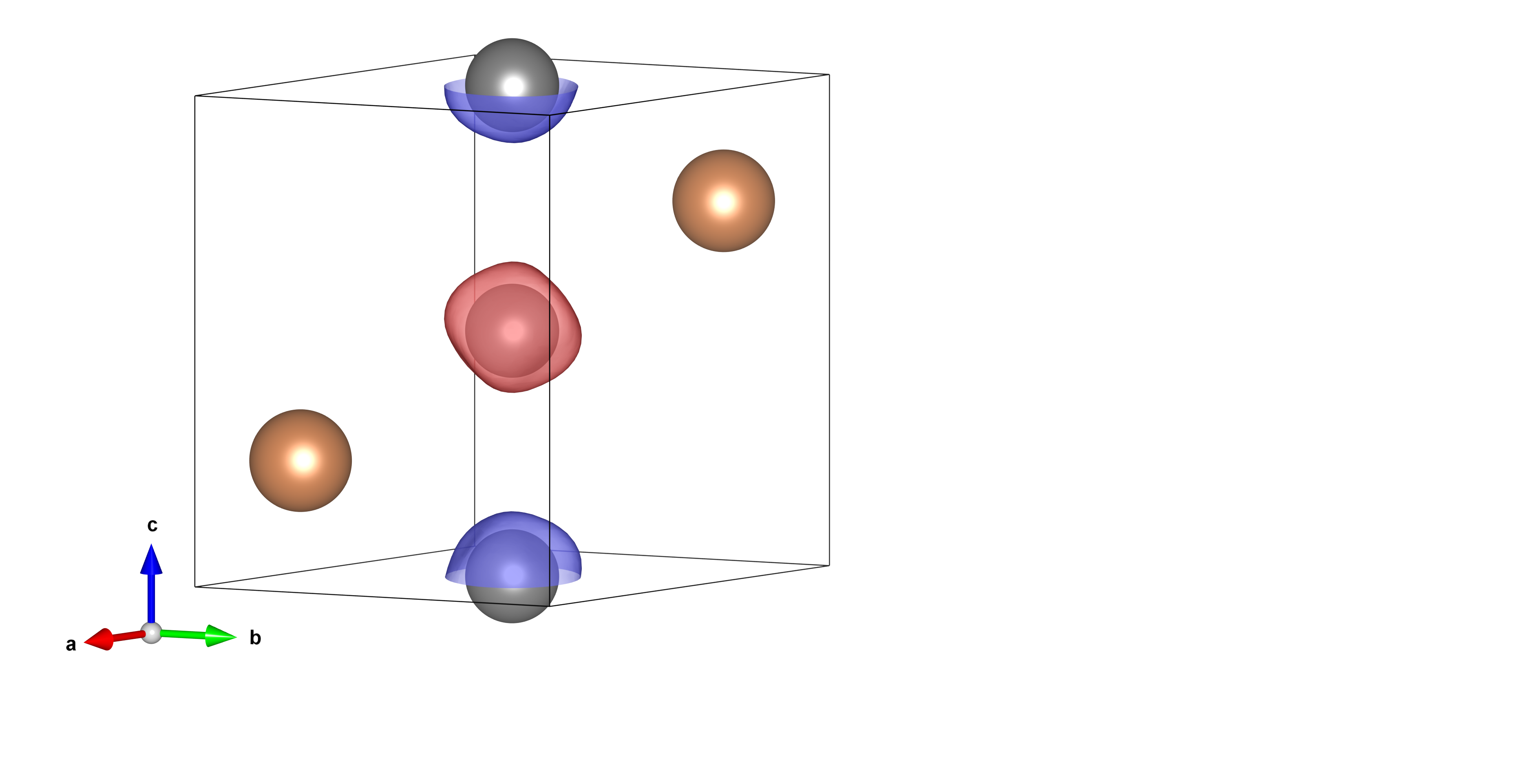}
    \caption{\label{fig:CrSb_spindens} Spin Density of CrSb projected along the $x$, $y$, and $z$ spin components (left to right). The noncollinear components $m_x$ and $m_y$ show the behavior of a magnetic octupole. Figure generated through the VESTA software~\cite{Momma2008Jun}.}
\end{figure}

\begin{figure}[H]
    \centering
    \includegraphics[width=0.95\linewidth]{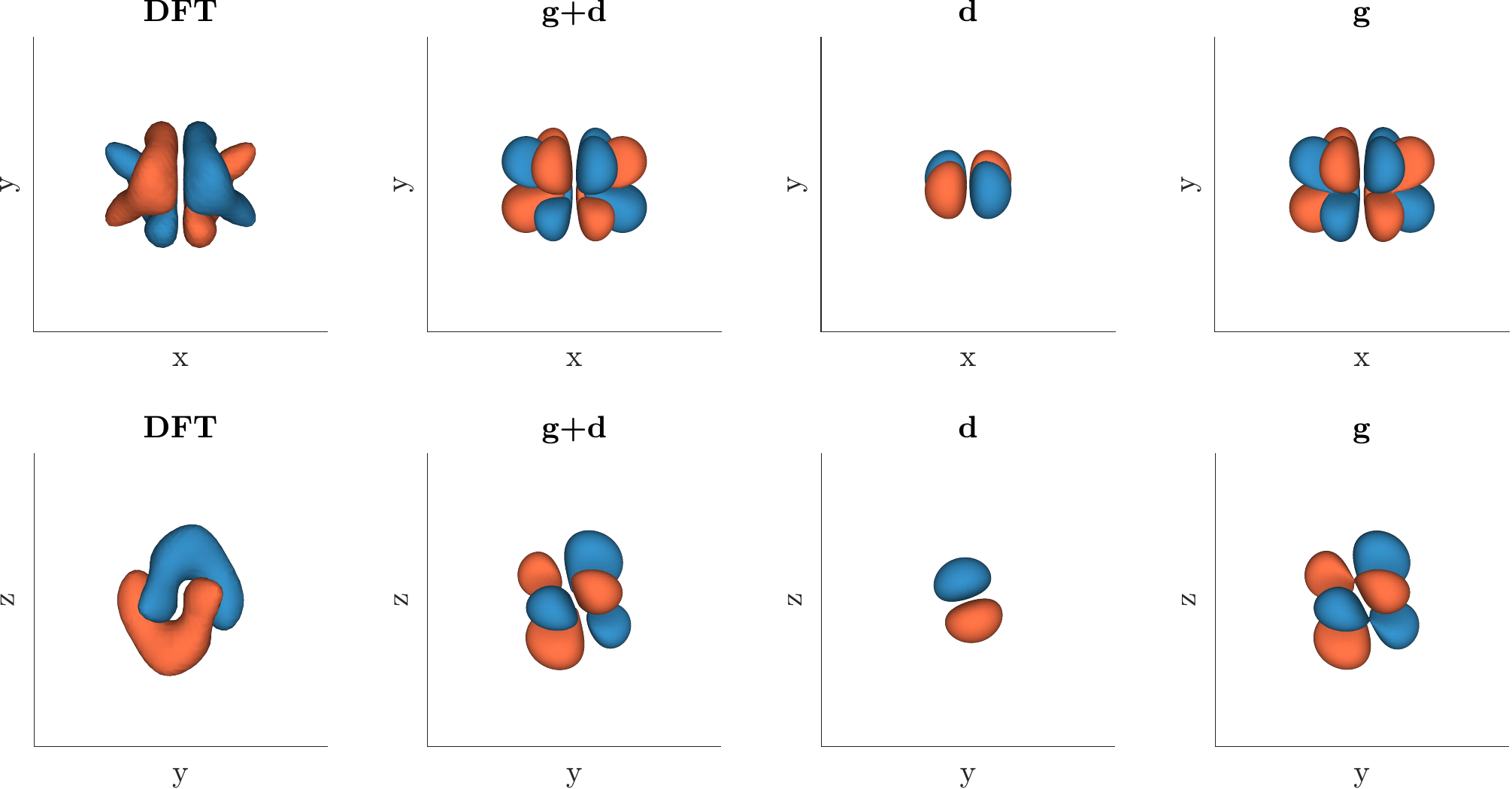}
    \caption{\label{fig:CrSb_mx_SumOfMultipoles} Noncollinear spin density $m_x$ of CrSb from DFT decomposed onto the tesseral harmonics.}
\end{figure}

\begin{figure}[H]
    \centering
    \includegraphics[width=0.33\linewidth, trim = 250 370 2200 0]{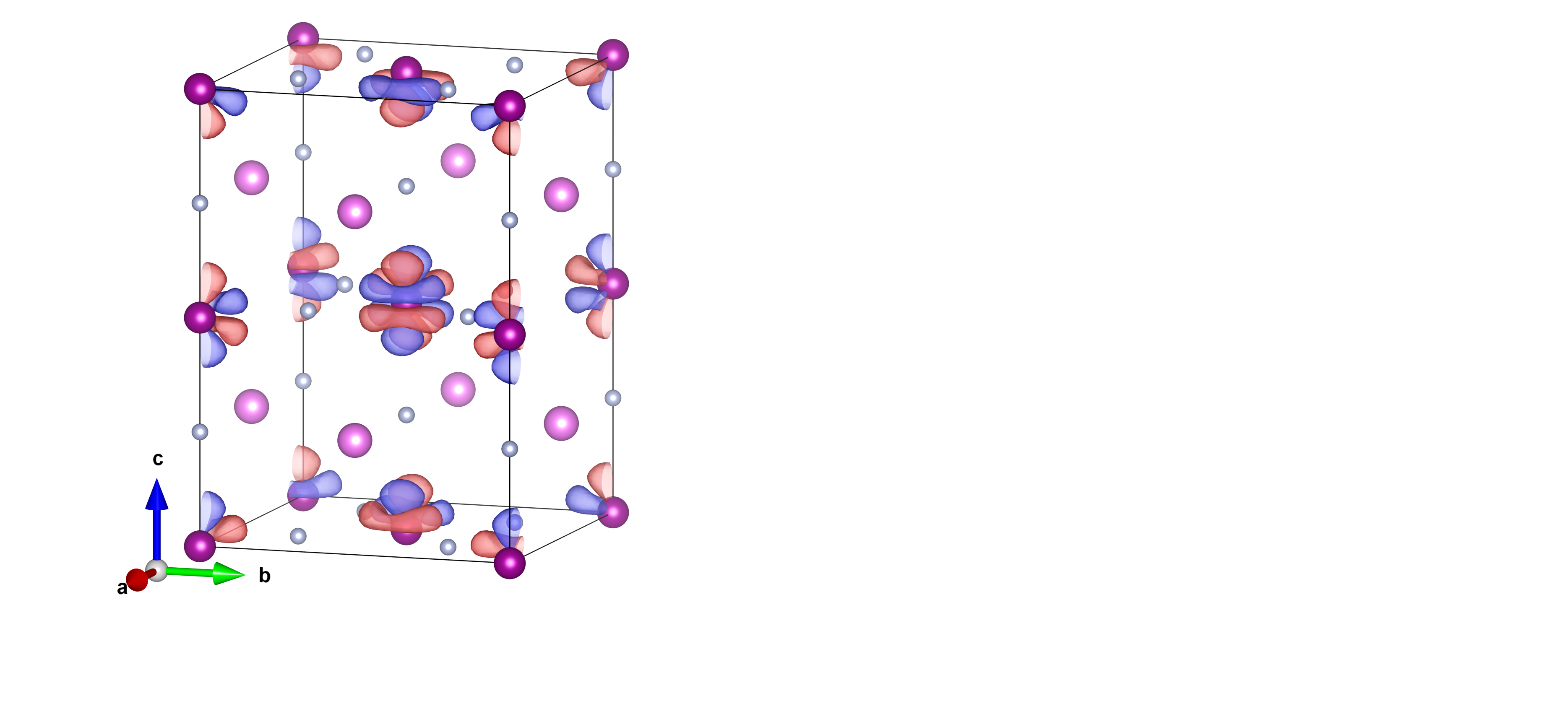}\includegraphics[width=0.33\linewidth, trim = 250 370 2200 0]{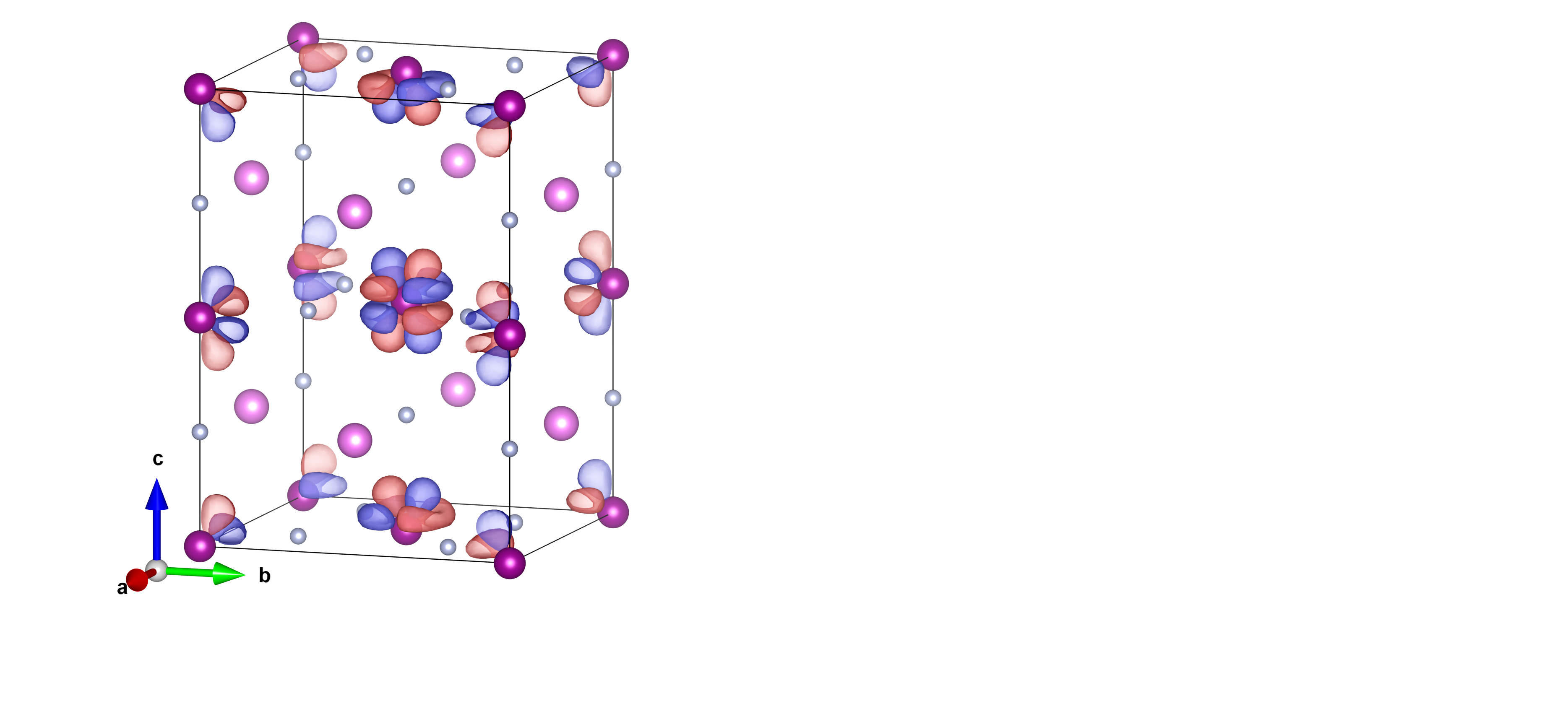}\includegraphics[width=0.33\linewidth, trim = 250 370 2200 0]{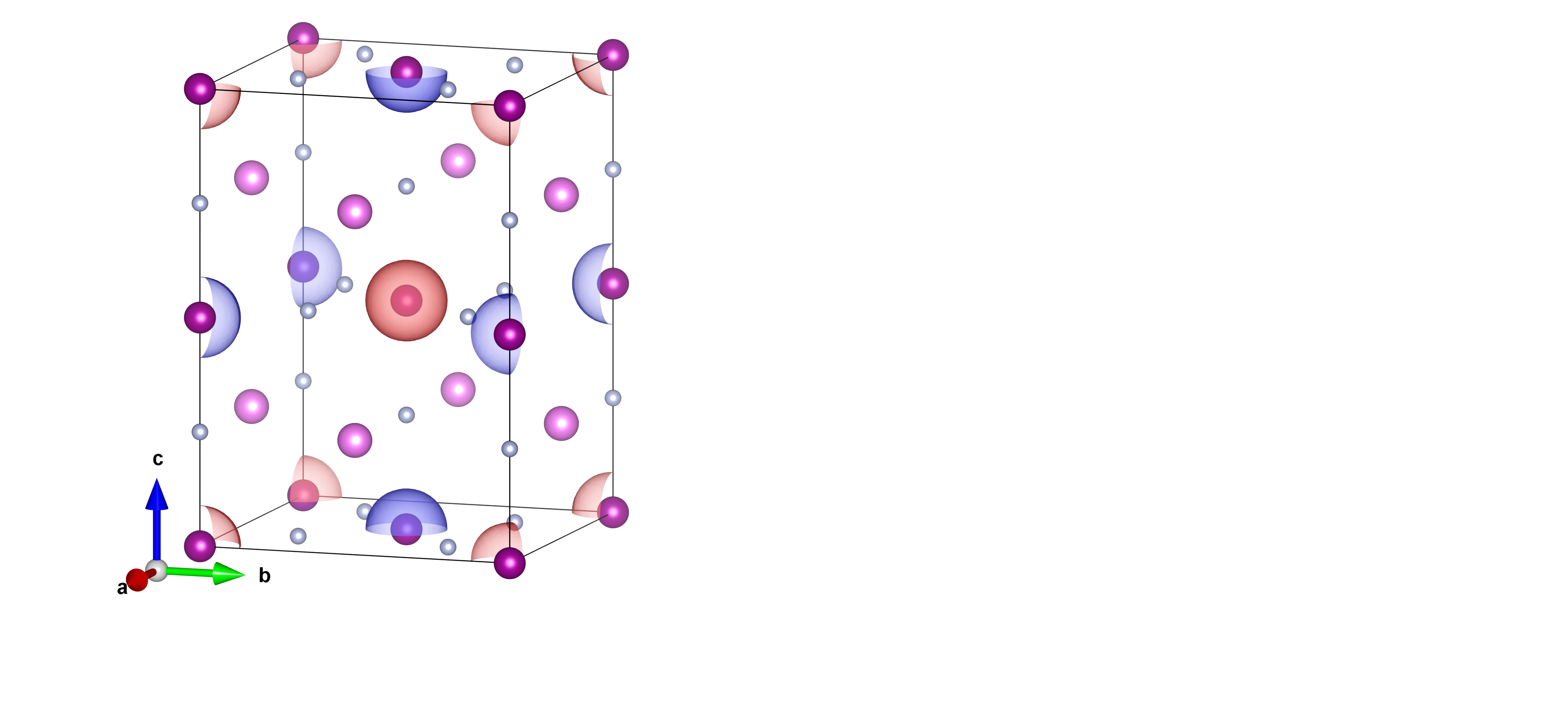}
    \caption{\label{fig:KMnF3_spindens} Spin Density of KMnF$_3$ from DFT projected along the $x$, $y$, and $z$ spin components (left to right). The noncollinear components $m_x$ and $m_y$ show the behavior of a magnetic dotriacontapole. Figure generated through the VESTA software~\cite{Momma2008Jun}.}
\end{figure}

\begin{figure}[H]
    \centering
    \includegraphics[width=0.95\linewidth]{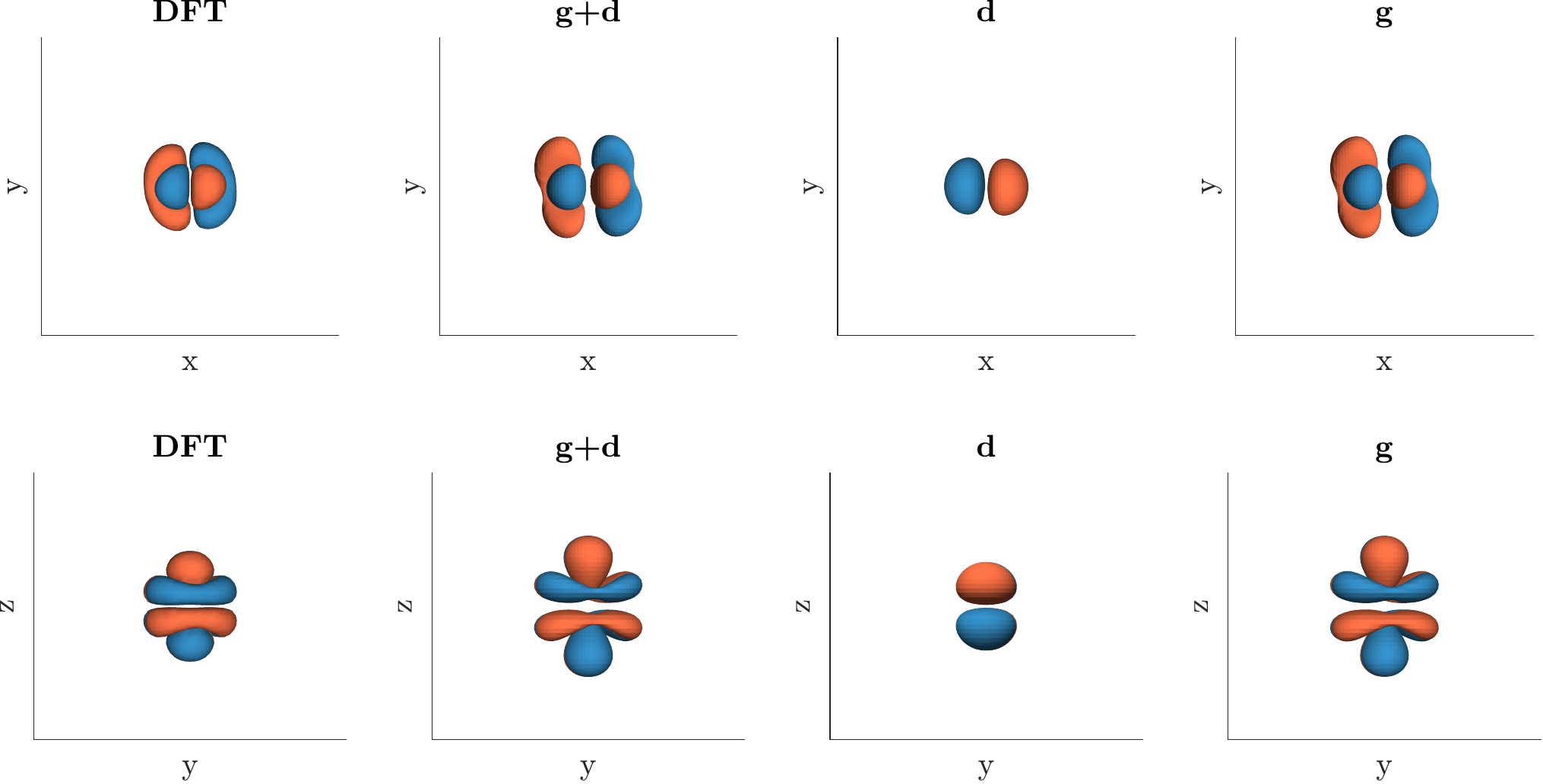}
    \caption{\label{fig:KMnF3_mx_SumOfMultipoles} Noncollinear spin density $m_x$ of KMnF$_3$ from DFT decomposed onto the tesseral harmonics.}
\end{figure}

From DFT we obtain the density matrix on a real space grid in the unit cell, which we show in Figures \ref{fig:MnF2_spindens}, \ref{fig:KMnF3_spindens}, \ref{fig:CrSb_spindens}. Using the spin density $m_i(\mathbf{r})$ we calculate the multipoles inside spheres $S_1, S_2$ centered on the magnetic atoms (notice that all three compounds have only two magnetic atoms in the primitive unit cell). Our choice for the size of the spheres is to use the atomic radii $R$ contained in the JTH PAW datasets, all reported in Tab.~\ref{tab:structures}. The magnitude of the magnetic multipoles is calculated by numerical integration, projecting the components of the spin density $m_i$ onto the tesseral harmonics $T_{\ell m}$,
\begin{equation}
    \alpha^{(1,2)}_{i,\ell m} = \frac{1}{V_{\rm uc}}\int_{S_1,S_2} d^3r\; m_i(\mathbf{r})\; T_{\ell m}(\mathbf{r})\;r^\ell.
\end{equation}
The tesseral harmonics we used are defined in Table \ref{tab:tess_harms}. All the coefficients that result from this projection are shown in Figures \ref{fig:MnF2_CubicHarmonics_SOx}, \ref{fig:MnF2_CubicHarmonics_SOy}, \ref{fig:MnF2_CubicHarmonics_SOz} for MnF$_2$, Figures \ref{fig:CrSb_CubicHarmonics_SOx}, \ref{fig:CrSb_CubicHarmonics_SOy}, \ref{fig:CrSb_CubicHarmonics_SOz} for CrSb and Figures \ref{fig:KMnF3_CubicHarmonics_SOx}, \ref{fig:KMnF3_CubicHarmonics_SOy}, \ref{fig:KMnF3_CubicHarmonics_SOz} for KMnF$_3$. Moreover, the coefficients $\alpha^{(1,2)}_{i,\ell m}$ have the atomic units of $\mu_Ba_0^\ell$, where $\mu_B$ is the Bohr magneton and $a_0$ is the Bohr length, but one has to be careful that their exact magnitude depends on details like the radius of integration $R$ and the normalization constant of the tesseral harmonics $T_{\ell m}$. Here, we chose to normalize the harmonics so that they are orthonormal on the surface of the unit sphere, up to a factor of $\sqrt{4\pi}$,  such that the lowest order $\ell=0$ is just the unity, $T_0=1$. This way, the projection of $m_i(\mathbf{r})$ onto $T_0$ simply gives the atomic dipole moment. In Figures \ref{fig:MnF2_mx_SumOfMultipoles}, \ref{fig:KMnF3_mx_SumOfMultipoles}, \ref{fig:CrSb_mx_SumOfMultipoles} we show how these coefficients can be used to reconstruct the angular behavior of the noncollinear spin density around the atoms. This reconstruction is only qualitative and not quantitatively exact; since the $T_{\ell m}$ are only a complete set on the surface of a sphere and not in all three-dimensional space, the inverse projection cannot be exact. The projection algorithm was implemented in the ProDenCeR code~\cite{Buiarelli2025}, which is available and open-source~\cite{drum}, along with the input files necessary to reproduce the results of this article.

\begin{table}[]
\begin{tabular}{l|l}
Name & Angular dependence \\
\hline
$T_0$ $(s)$ & $1$ \\ \hline
$T_2$ $(d)$ & $d_{xy} = \sqrt{\frac{15}{4}}\,2xy/r^2$ \\
  & $d_{yz} = \sqrt{\frac{15}{4}}\,2yz/r^2$ \\
  & $d_{z^2} = \sqrt{\frac{5}{4}}\,(3z^2-r^2)/r^2$ \\
  & $d_{xz} = \sqrt{\frac{15}{4}}\,2xz/r^2$ \\
  & $d_{x^2-y^2} = \sqrt{\frac{15}{4}}\,(x^2-y^2)/r^2$ \\ \hline
$T_4$ $(g)$ & $g_{xy(x^2-y^2)} = \frac{3}{4}\sqrt{35}\,xy(x^2-y^2)/r^4$ \\
  & $g_{xy(x^2-y^2)} = \frac{3}{2}\sqrt{35}\,xy(x^2-y^2)/r^4$ \\
  & $g_{yz(3x^2-y^2)} = \frac{3}{4}\sqrt{70}\,yz(3x^2-y^2)/r^4$ \\
  & $g_{xyz^2} = \frac{3}{2}\sqrt{5}\,xy(7z^2-r^2)/r^4$ \\
  & $g_{yz^3} = \frac{3}{8}\sqrt{5}\,yz(7z^2-3r^2)/r^4$ \\
  & $g_{z^4} = \frac{3}{8}\,(35z^4-30z^2r^2+3r^4)/r^4$ \\
  & $g_{xz^3} = \frac{3}{8}\sqrt{5}\,xz(7z^2-3r^2)/r^4$ \\
  & $g_{(x^2-y^2)z^2} = \frac{3}{4}\sqrt{5}\,(x^2-y^2)(7z^2-r^2)/r^4$ \\
  & $g_{xz(x^2-3y^2)} = \frac{3}{4}\sqrt{70}\,xz(x^2-3y^2)/r^4$ \\
  & $g_{x^2y^2} = \frac{3}{8}\sqrt{35}\,(6x^2y^2-x^4-y^4)/r^4$ \\
\end{tabular}
\caption{Inversion even tesseral harmonics $T_{\ell=0,2,4}$. In the text they are also referred to as electric multipoles.}
\label{tab:tess_harms}
\end{table}

\newpage
\section{Group theoretical analysis and DFT results}

\subsection{Rutile MnF$_2$}

\begin{table}[]
\begin{tabular}{c|ccc|c|c}
         & $\rm{m}_{001}$ & $\rm{m}_{1\bar{1}0}$ & $\rm{m}_{110}$ & $m_{\ell=1}$ & $T_{\ell=0,2,4}$   \\ \hline
$A_g$    & \;1         & \;1         & \;1         &              & 1, $d_{xy}$, $d_{z^2}$, $g_{xyz^2}$, $g_{x^2y^2}$, $g_{z^4}$      \\
$B_{1g}$ & \;1         & -1        & -1        & $m_z$        & $d_{x^2-y^2}$, $g_{xy(x^2-y^2)}$, $g_{(x^2-y^2)z^2}$  \\
$B_{2g}$ & -1        & \;1         & -1        & $m_x-m_y$    & $d_{xz}+d_{yz}$, $g_{xz^3}+g_{yz^3}$, $g_{xz(x^2-3y^2)}+d_{yz(y^2-3x^2)}$\\
$B_{3g}$ & -1        & -1        & \;1         & $m_x+m_y$    & $d_{xz}-d_{yz}$, $g_{xz^3}-g_{yz^3}$, $g_{xz(x^2-3y^2)}-g_{yz(y^2-3x^2)}$
\end{tabular}
\caption{Character table for $mmm$ ($D_{2h}$), the site symmetry group of the Mn ions in MnF$_2$. Only the inversion even irreducible representations are shown.}
\label{tab:mmm_table}
\end{table}

\begin{table}[]
\begin{tabular}{c|ccc}
              & $\left\{4_{001}|\frac{1}{2},\frac{1}{2},\frac{1}{2}\right\}$ & $\left\{\rm{m}_{100}|\frac{1}{2},\frac{1}{2},\frac{1}{2}\right\}$ & $\left\{\rm{m}_{110}|0,0,0\right\}$ \\ \hline
$m\Gamma_1^+$ & \;1                       & \;1                       & \;1                 \\
$m\Gamma_2^+$ & -1                      & \;1                       & -1                \\
$m\Gamma_3^+$ & \;1                       & -1                      & -1                \\
$m\Gamma_4^+$ & -1                      & -1                      & \;1                
\end{tabular}
\caption{Character table for $P4_2/mnm.1'$, the paramagnetic space group for the rutile MnF$_2$. Only the inversion even, time reversal odd, irreducible representations are shown.}
\label{tab:P42mnm_table}
\end{table}

The paramagnetic rutile structure of MnF$_2$ belongs to the magnetic Space Group (mSG) $P4_2/mnm.1'$ (\#136.496). The $\Gamma$-point irreps of $P4_2/mnm.1'$ are shown in Table \ref{tab:P42mnm_table}. The magnetic configuration that has antiparallel magnetic moments on the two Mn ions results from the condensation of the $m\Gamma_2^+$ irrep and yields the mSG $P4_2'/mnm'$ (\#136.499). The Mn ions are at the Wyckoff position $2a$ which has site symmetry $mmm$ ($D_{2h}$), for which we show the character table in Table \ref{tab:mmm_table}. Through the Bilbao Crystallographic Server~\cite{Elcoro2017Oct} we check that the site symmetry irrep that induces $m\Gamma_2^+$ in the unit cell is $B_{1g}$. In Table \ref{tab:mmm_table} we also show the symmetry properties of magnetic dipoles and electric quadrupoles and hexadecapoles. By combining these, we can get magnetic octupoles and dotriacontapoles. Consequently we check which magnetic multipoles, up to $\ell=5$, belong to the $B_{1g}$ irrep and claim that those are allowed around the magnetic ions in the magnetic phase. Regarding the collinear multipoles, the ones involving only $m_z$, the magnetic multipoles that transform like $B_{1g}$ are given by multiplying $m_z$ with an $A_g$ electric multipole $T_{\ell=0,2,4}$, yielding:

\begin{equation}
    m_{\ell=1,3,5}=\left\{\begin{matrix}
        d_{xy}m_z,\; g_{xyz^2}m_z, & \textrm{odd under } 4_{001} \rightarrow m\Gamma_2^+,\\
        m_z,\; d_{z^2}m_z,\; g_{x^2y^2}m_z,\; g_{z^4}m_z, & \textrm{even under } 4_{001} \rightarrow m\Gamma_3^+.
\end{matrix}\right. 
\end{equation}

We have classified the $B_{1g}$ magnetic multipoles depending on how they behave under the 4-fold rotation, which also classifies them as the $P4_2/mnm.1'$ irreps $m\Gamma_2^+$ or $m\Gamma_3^+$. As a space group operation, the 4-fold rotation is followed by a $(1/2,1/2,1/2)$ non-symmorphic translation that connects the two Mn ions with opposite magnetic moment. The 4-fold even multipoles cannot be in-phase on the two ions (this is clear for the dipole $m_z$), otherwise it would lead to a ferromagnetic phase. For this reason, we introduce the Pauli matrices $\tau_0\in m\Gamma_1^+$ and $\tau_z\in m\Gamma_3^+$ in sublattice space, where $\tau_0$ represents an object being in-phase on the two sites and $\tau_z$, which is odd under all the non-symmorphic operations and even under the symmorphic ones, represents an object being out-of-phase on the two sites. In this way, because $m\Gamma_3^+\times m\Gamma_4^+ = m\Gamma_2^+$, the collinear magnetic multipoles on the Mn ions are

\begin{equation}
    m\Gamma_2^+\; : \; m_{\ell=1,3,5}=\left\{\begin{matrix}
        \left\{d_{xy},\; g_{xyz^2} \right\} m_z\tau_0, \\
        \left\{1,\; d_{z^2},\; g_{x^2y^2},\; g_{z^4} \right\} m_z\tau_z.
\end{matrix}\right. 
\end{equation}

A similar logic can be applied to the noncollinear magnetic multipoles. To get the multipoles that have site symmetry $B_{1g}$, we need to take the products $B_{2g}\times B_{3g}$ and then classify them by how they transform under the 4-fold rotation. We then impose that the ones that are even under rotation must be out-of-phase on the two ions in order to give the mSG irrep $m\Gamma_2^+$, which ultimately yields

\begin{equation}
    m\Gamma_2^+\; : \; m_{\ell=3,5}=\left\{\begin{matrix}
        \left\{d_{yz},\; g_{yz^3},\; g_{yz(y^2-3x^2)} \right\} m_x\tau_0 + \left\{d_{xz},\; g_{xz^3},\; g_{xz(x^2-3y^2)} \right\} m_y\tau_0, \\
        \left\{d_{xz},\; g_{xz^3},\; g_{xz(x^2-3y^2)} \right\} m_x\tau_z + \left\{d_{yz},\; g_{yz^3},\; g_{yz(y^2-3x^2)} \right\} m_y\tau_z.
\end{matrix}\right. 
\end{equation}

\begin{figure}[H]
    \centering
    \includegraphics[width=\linewidth]{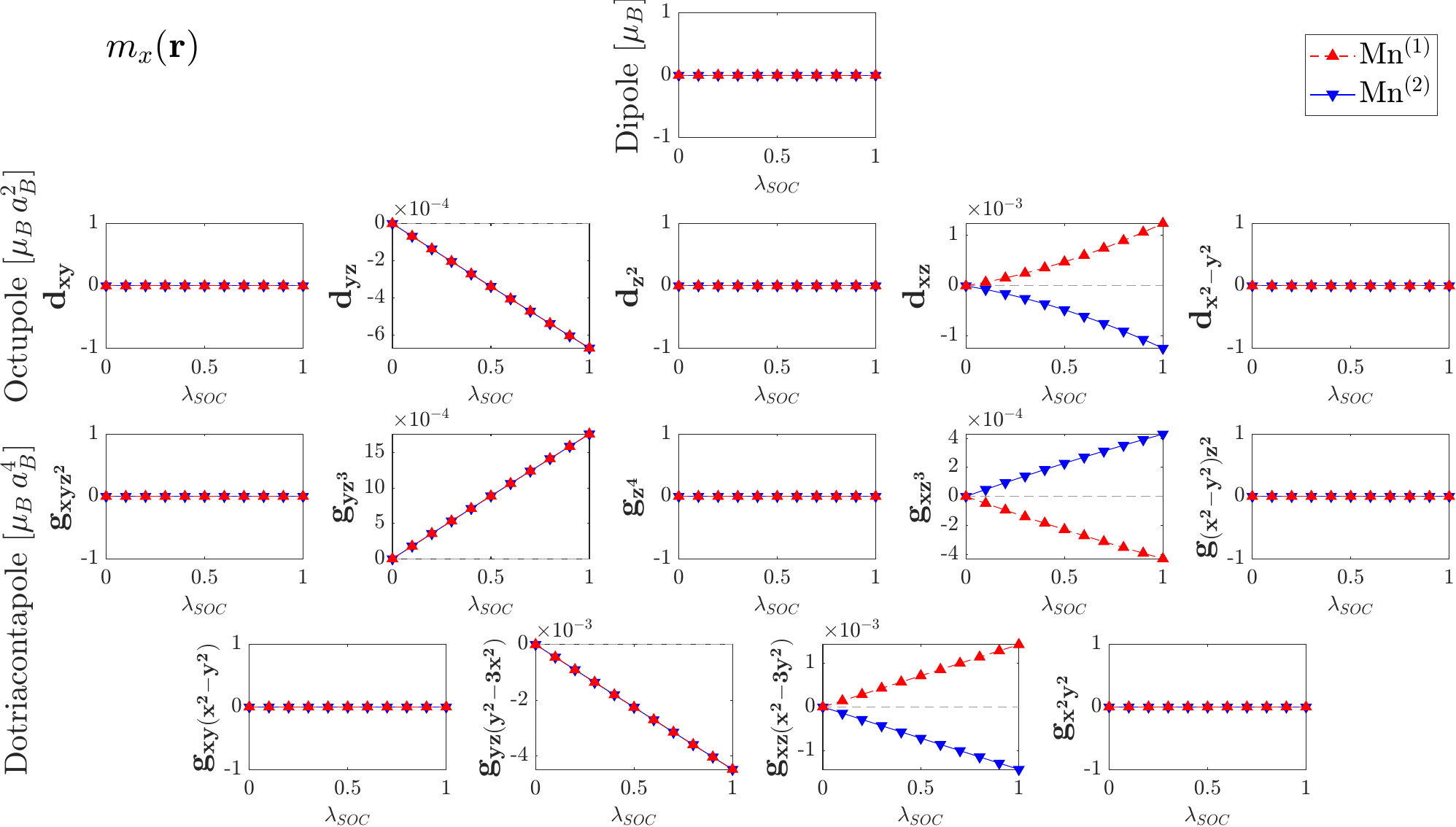}
    \caption{\label{fig:MnF2_CubicHarmonics_SOx} Multipole coefficients $\alpha^{(1,2)}_{x,\ell m}$ obtained by projecting the $m_x(\mathbf{r})$ component of the spin density in MnF$_2$ onto the inversion-even tesseral harmonics $T_{\ell m}$ up to $\ell=4$, as a function of spin-orbit coupling $\lambda_{SOC}$.}
\end{figure}

\begin{figure}[H]
    \centering
    \includegraphics[width=\linewidth]{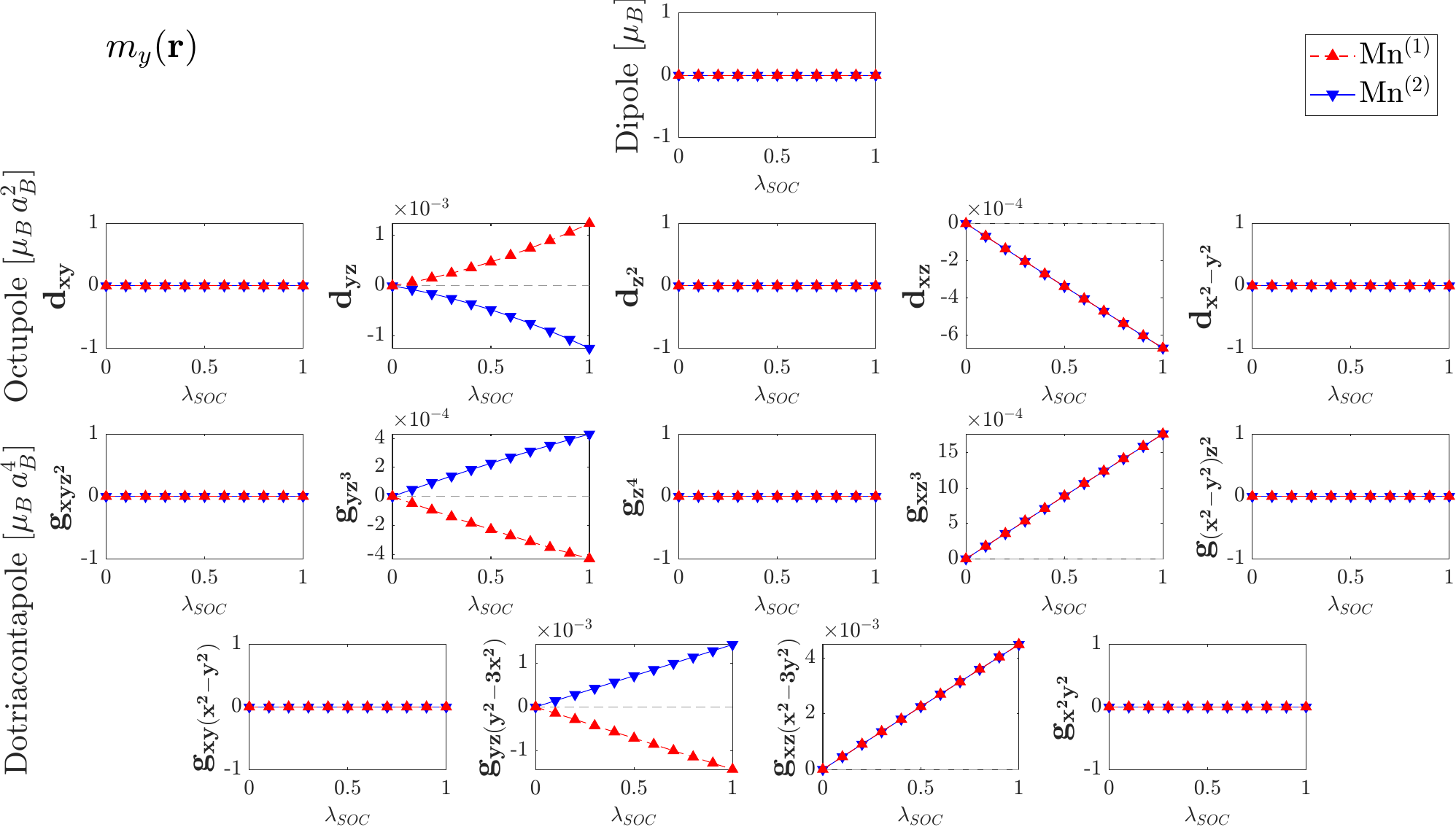}
    \caption{\label{fig:MnF2_CubicHarmonics_SOy} Multipole coefficients $\alpha^{(1,2)}_{y,\ell m}$ obtained by projecting the $m_y(\mathbf{r})$ component of the spin density in MnF$_2$ onto the inversion-even tesseral harmonics $T_{\ell m}$ up to $\ell=4$, as a function of spin-orbit coupling $\lambda_{SOC}$.}
\end{figure}

\begin{figure}[H]
    \centering
    \includegraphics[width=\linewidth]{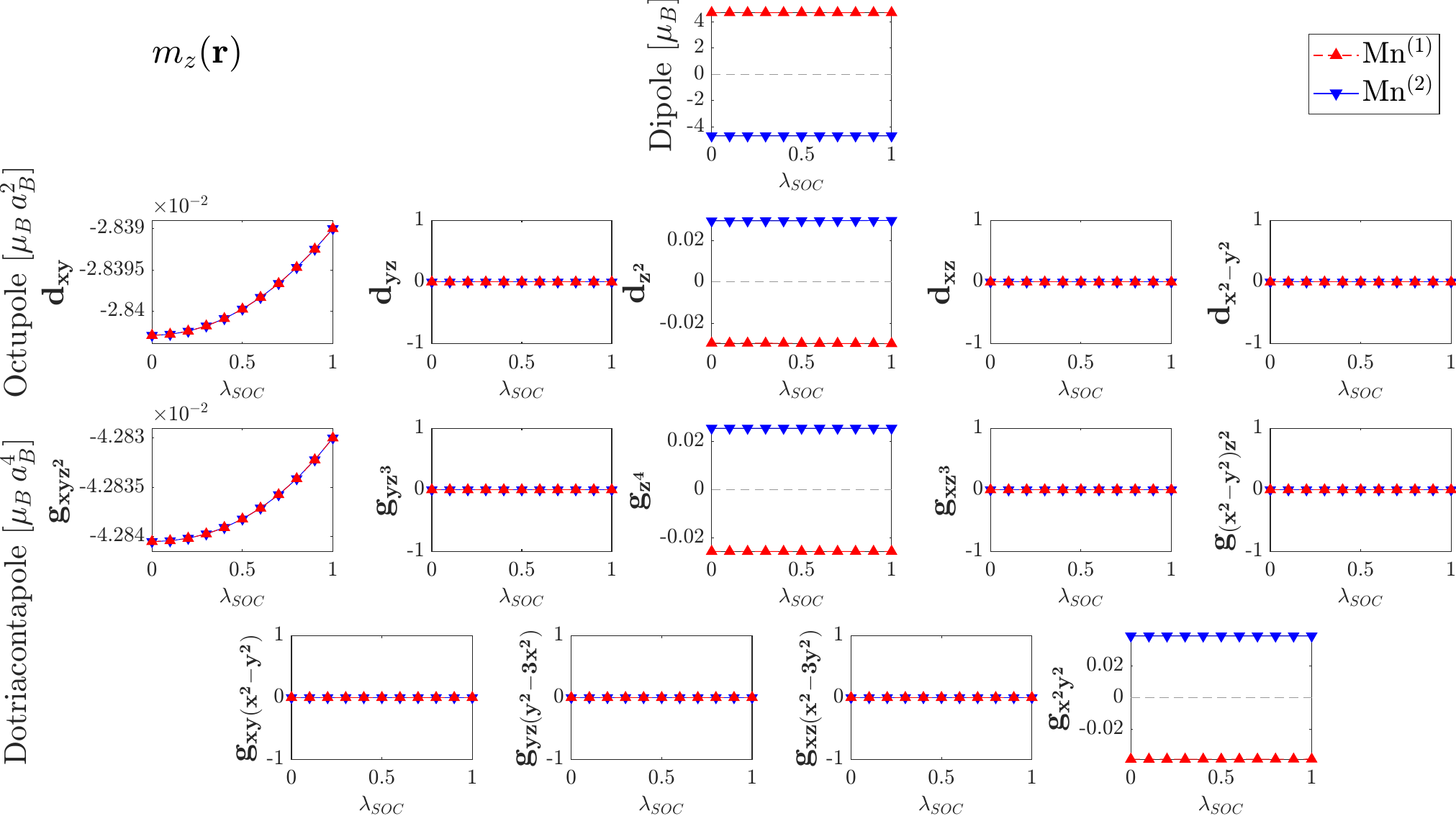}
    \caption{\label{fig:MnF2_CubicHarmonics_SOz} Multipole coefficients $\alpha^{(1,2)}_{z,\ell m}$ obtained by projecting the $m_z(\mathbf{r})$ component of the spin density in MnF$_2$ onto the inversion-even tesseral harmonics $T_{\ell m}$ up to $\ell=4$, as a function of spin-orbit coupling $\lambda_{SOC}$.}
\end{figure}

These results are confirmed by our DFT calculations. As can be seen in Figures \ref{fig:MnF2_CubicHarmonics_SOx}, \ref{fig:MnF2_CubicHarmonics_SOy}, \ref{fig:MnF2_CubicHarmonics_SOz}, the multipoles that are in-phase have the same value on both atoms, while the out-of-phase multipoles have same magnitude but opposite sign on the two atoms. The lowest order collinear magnetic multipole that is in-phase is $d_{xy}m_z$, which is consistent with the typical observation that altermagnetism in MnF$_2$ is of the $d$-wave kind. We find that there should also be a weak altermagnetism of the $d$-wave kind in the noncollinear components, since $d_{yz}m_x+d_{xz}m_y$ is in-phase. Moreover, these $d$-wave components correspond to the piezomangetic tensor components that are allowed for the mSG: $\varepsilon_{xy}$ will lead to a weak ferromagnetic moment $m_z$, $\varepsilon_{yz}$ will lead to a weak ferromagnetic moment $m_x$ and $\varepsilon_{xz}$ will lead to a weak ferromagnetic moment $m_y$. These results can be generalized to any other tensor with Jahn symbol $aeV[V^2]$ and not just the piezomagnetic one.

\newpage
\subsection{Hexagonal CrSb}

\begin{table}[]
\begin{tabular}{c|cc|c|c}
         & $3_{001}$ & $\rm{m}_{100}$ & $m_{\ell=1}$ & $T_{\ell=0,2,4}$   \\ \hline
$A_{1g}$ & \;1         & \;1         &              & 1, $d_{z^2}$, $g_{yz(y^2-3x^2)}$, $g_{z^4}$      \\
$A_{2g}$ & \;1         & -1        & $m_z$        & $g_{xz(x^2-3y^2)}$  \\
$E_{g}$  & -1        & \;0         & $\left\{m_x, m_y\right\}$    & $\left\{d_{xy}, d_{x^2-y^2}\right\}$, $\left\{d_{xz}, d_{yz}\right\}$, $\left\{g_{xy(x^2-y^2)}, g_{x^2y^2}\right\}$, $\left\{g_{xz^3}, g_{yz^3}\right\}$, $\left\{g_{xyz^2}, g_{(x^2-y^2)z^2}\right\}$
\end{tabular}
\caption{Character table for $\bar{3}m$ ($D_{3d}$), the site symmetry group of the Cr ions in CrSb. Only the inversion even irreducible representations are shown. All the functions are given in cartesian, not primitive, coordinates.}
\label{tab:3-m_table}
\end{table}

The paramagnetic structure of CrSb belongs to the magnetic Space Group (mSG) $P3_4/mmc.1'$ (\#194.264). The irreps of $P6_3/mmc.1'$ are shown in Table \ref{tab:P63mmc_table}. The magnetic configuration that has antiparallel magnetic moments on the two Cr ions comes from the condensation of the $m\Gamma_4^+$ irrep and yields the mSG $P6_3'/m'm'c$ (\#194.268). The Cr ions are at the Wyckoff position $2a$ which has site symmetry $\bar{3}m$ ($D_{3d}$), for which we show the character table in Table \ref{tab:3-m_table}. The site symmetry irrep that induces $m\Gamma_4^+$ in the unit cell is $A_{2g}$. In Table \ref{tab:3-m_table} we also show the symmetry properties of magnetic dipoles and electric quadrupoles and hexadecapoles. By combining these, we can get magnetic octupoles and dotriacontapoles. Regarding the collinear multipoles, the ones involving only $m_z$, the magnetic multipoles that transform like $A_{2g}$ are given by multiplying $m_z$ with an $A_{1g}$ electric multipole $T_{\ell=2,4}$, yielding:

\begin{equation}
    m_{\ell=1,3,5}=\left\{\begin{matrix}
        g_{yz(y^2-3x^2)}m_z, & \textrm{odd under } 6_{001} \rightarrow m\Gamma_4^+,\\
        m_z,\; d_{z^2}m_z,\; g_{z^4}m_z, & \textrm{even under } 6_{001} \rightarrow m\Gamma_2^+.
\end{matrix}\right. 
\end{equation}

We have classified the $A_{2g}$ magnetic multipoles depending on how they behave under the 6-fold rotation. As a space group operation, the 6-fold rotation is followed by a $(0,0,1/2)$ non-symmorphic translation that connects the two Cr ions with opposite magnetic moment. Similarly as in the rutile case, the 6-fold odd multipoles are in-phase while the 6-fold even multipoles must be out-of-phase to yield the correct magnetic space group, thus
\begin{equation}
    m\Gamma_4^+\; : \; m_{\ell=1,3,5}=\left\{\begin{matrix}
        g_{yz(y^2-3x^2)} m_z\tau_0, \\
        \left\{1,\; d_{z^2},\; g_{z^4} \right\} m_z\tau_z,
\end{matrix}\right. 
\end{equation}
where $\tau_0\in m\Gamma_1^+$ and $\tau_z\in m\Gamma_3^+$. We repeat the same operations for the noncollinear components. This time, by combining $\left\{m_x, m_y\right\}$ and the $E_g$ electric quadrupoles and hexadecapoles we get $E_g\times E_g = A_{1g} \oplus A_{2g} \oplus E_g$, where we are interested in the $A_{2g}$ contribution only

\begin{equation}
    m\Gamma_4^+\; : \; m_{\ell=3,5}=\left\{\begin{matrix}
        \left\{d_{yz},\; g_{yz^3} \right\} m_x\tau_0 + \left\{d_{xz},\; g_{xz^3} \right\} m_y\tau_0, \\
        \left\{d_{xy},\; g_{xyz^2},\; g_{xy(x^2-y^2)} \right\} m_x\tau_z + \left\{d_{x^2-y^2},\; g_{(x^2-y^2)z^2},\; g_{x^2y^2} \right\} m_y\tau_z.
\end{matrix}\right. 
\end{equation}

\begin{figure}[H]
    \centering
    \includegraphics[width=\linewidth]{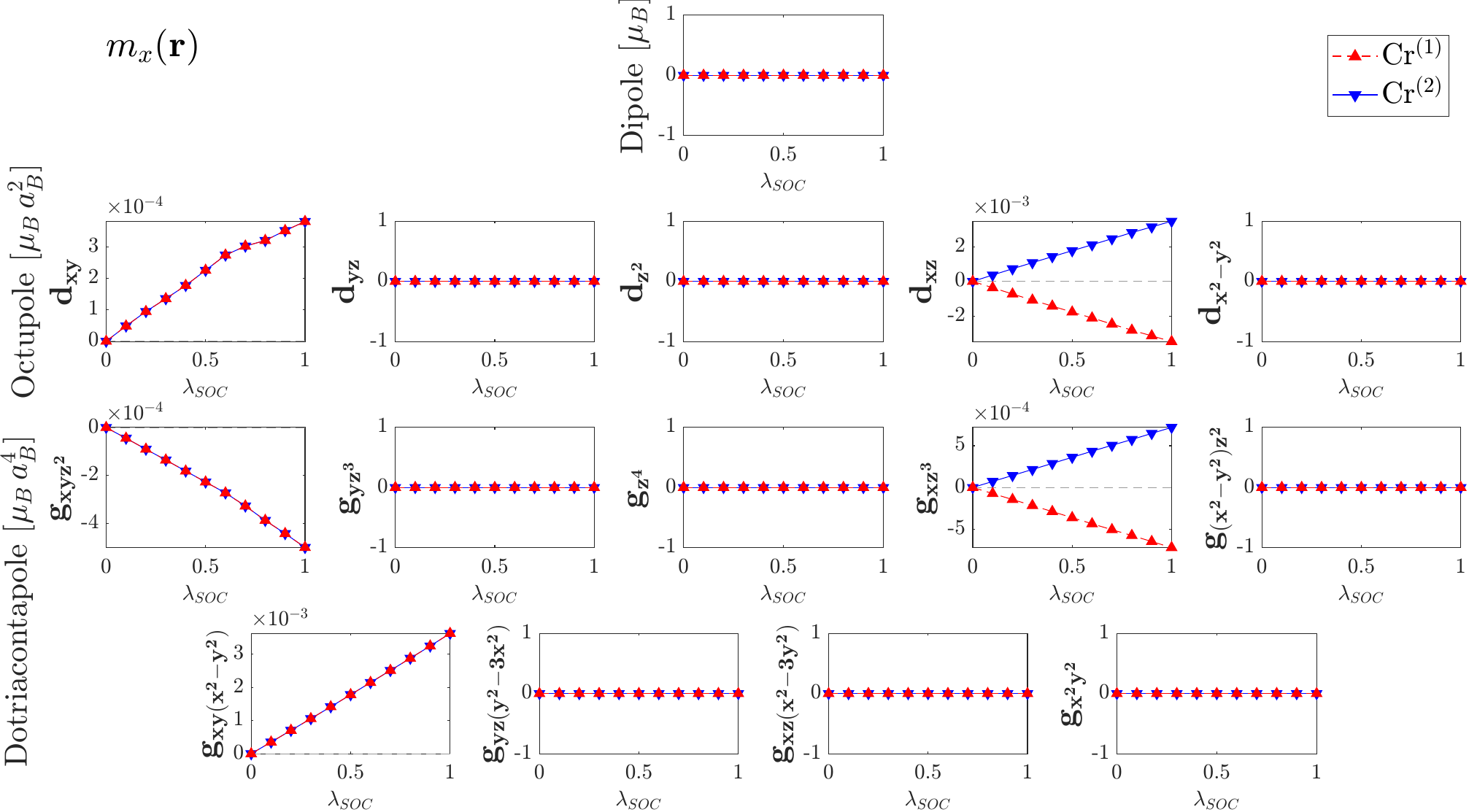}
    \caption{\label{fig:CrSb_CubicHarmonics_SOx} Multipole coefficients $\alpha^{(1,2)}_{x,\ell m}$ obtained by projecting the $m_x(\mathbf{r})$ component of the spin density in CrSb onto the inversion-even tesseral harmonics $T_{\ell m}$ up to $\ell=4$, as a function of spin-orbit coupling $\lambda_{SOC}$.}
\end{figure}

\begin{figure}[H]
    \centering
    \includegraphics[width=\linewidth]{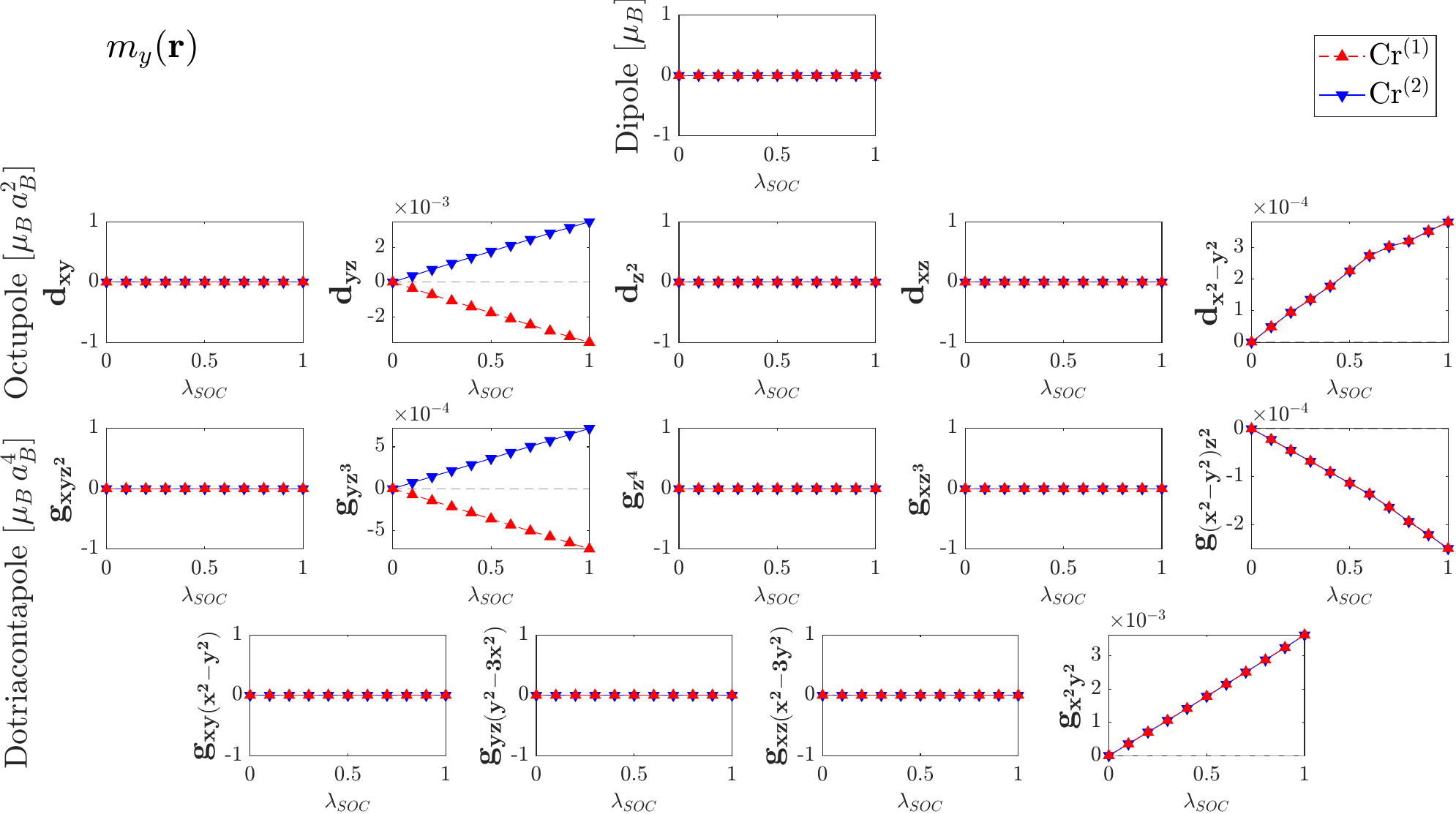}
    \caption{\label{fig:CrSb_CubicHarmonics_SOy} Multipole coefficients $\alpha^{(1,2)}_{y,\ell m}$ obtained by projecting the $m_y(\mathbf{r})$ component of the spin density in CrSb onto the inversion-even tesseral harmonics $T_{\ell m}$ up to $\ell=4$, as a function of spin-orbit coupling $\lambda_{SOC}$.}
\end{figure}

\begin{figure}[H]
    \centering
    \includegraphics[width=\linewidth]{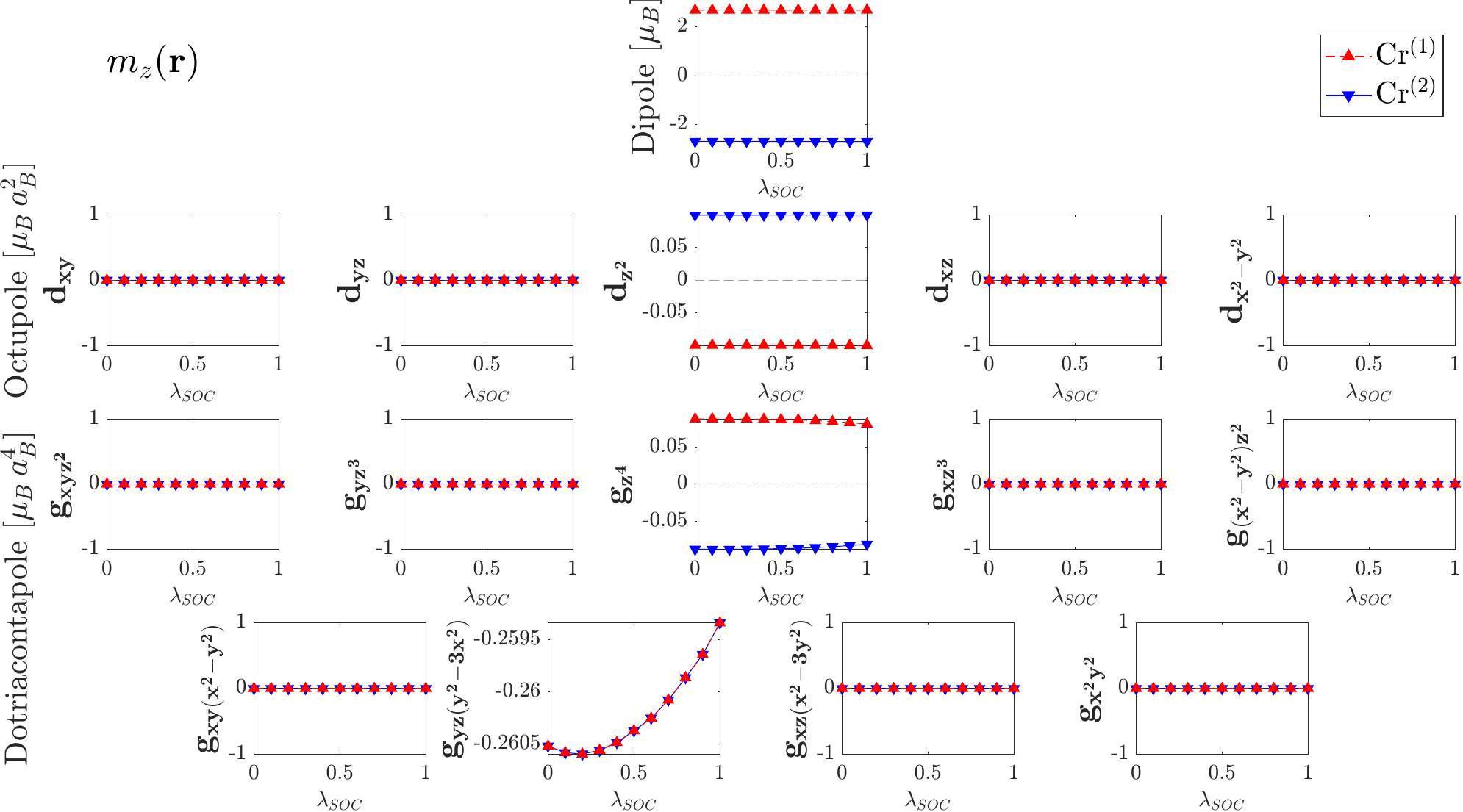}
    \caption{\label{fig:CrSb_CubicHarmonics_SOz} Multipole coefficients $\alpha^{(1,2)}_{z,\ell m}$ obtained by projecting the $m_z(\mathbf{r})$ component of the spin density in CrSb onto the inversion-even tesseral harmonics $T_{\ell m}$ up to $\ell=4$, as a function of spin-orbit coupling $\lambda_{SOC}$.}
\end{figure}

Again, this analysis is confirmed by our DFT results in Figures \ref{fig:CrSb_CubicHarmonics_SOx}, \ref{fig:CrSb_CubicHarmonics_SOy} and \ref{fig:CrSb_CubicHarmonics_SOz}. For the collinear component $m_z$ the lowest order multipole that is in-phase is $g_{yz(y^2-3x^2)}m_z$, thus giving g-wave altermagnetism. For the noncollinear components, we find d-wave altermagnetism instead, having the lowest order in-phase multipole $d_{xy}m_x + d_{x^2-y^2}m_y$. Regarding piezomagnetism, this is in agreement with the expected selection rules for the piezomagnetic tensor for this mSG: no strain can lead to a weak ferromagnetic moment $m_z$, $\varepsilon_{xy}$ leads to a weak ferromagnetic moment $m_x$ and $\varepsilon_{xx}-\varepsilon_{yy}$ leads to a weak ferromagnetic moment $m_y$. These results can be generalized to any other tensor with Jahn symbol $aeV[V^2]$ and not just the piezomagnetic one.

\begin{table}[]
\begin{tabular}{c|ccc}
              & $\left\{6_{001}|0,0,\frac{1}{2}\right\}$ & $\left\{\rm{m}_{120}|0,0,\frac{1}{2}\right\}$ & $\left\{\rm{m}_{100}|0,0,0\right\}$ \\ \hline
$m\Gamma_1^+$ & \;1                       & \;1                       & \;1                 \\
$m\Gamma_2^+$ & \;1                       & -1                      & -1                \\
$m\Gamma_3^+$ & -1                      & -1                      & \;1                 \\
$m\Gamma_4^+$ & -1                      & \;1                       & -1                
\end{tabular}
\caption{Character table for $P6_3/mmc.1'$, the paramagnetic space group for the rutile CrSb. Only the inversion even, time reversal odd, irreducible representations are shown.}
\label{tab:P63mmc_table}
\end{table}

\newpage
\subsection{Perovskite KMnF$_3$}

\begin{table}[]
\begin{tabular}{c|cc|c|c}
         & $4_{001}$ & $\rm{m}_{001}$ & $m_{\ell=1}$ & $T_{\ell=0,2,4}$   \\ \hline
$A_{g}$ & \;1         & \;1         & $m_z$              & 1, $d_{z^2}$, $g_{xy(x^2-y^2)}$, $g_{z^4}$, $g_{x^2y^2}$      \\ 
$B_{g}$ & -1         & \;1        &        & $d_{xy}$, $d_{x^2-y^2}$, $g_{xyz^2}$, $g_{(x^2-y^2)z^2}$ \\
$E_{g}$  & \;0        & -2         & $\left\{m_x, m_y\right\}$    & $\left\{d_{xz}, d_{yz}\right\}$, $\left\{g_{xz^3}, g_{yz^3}\right\}$
\end{tabular}
\caption{Character table for $4/m$ ($C_{4h}$), the site symmetry group of the Mn ions in tetragonal KMnF$_3$. Only the inversion even irreducible representations are shown.}
\label{tab:4m_table}
\end{table}

\begin{table}[]
\begin{tabular}{c|ccc}
              & $\left\{4_{001}|0,0,0\right\}$ & $\left\{\rm{m}_{100}|0,0,\frac{1}{2}\right\}$ & $\left\{\rm{m}_{110}|0,0,\frac{1}{2}\right\}$ \\ \hline
$m\Gamma_1^+$ & \;1                       & \;1                       & \;1                 \\
$m\Gamma_2^+$ & -1                      & \;1                       & -1                \\
$m\Gamma_3^+$ & \;1                       & -1                      & -1                \\
$m\Gamma_4^+$ & -1                      & -1                      & \;1                
\end{tabular}
\caption{Character table for $I4/mcm.1'$, the paramagnetic space group for the tetragonal perovskite KMnF$_3$. Only the inversion even, time reversal odd, irreducible representations are shown.}
\label{tab:I4mcm_table}
\end{table}

Without any magnetic moments present, tetragonal perovskite KMnF$_3$ would belong to the mSG $I4/mcm.1'$ (\#140.541). The irreps of $I4/mcm.1'$ are shown in Table \ref{tab:I4mcm_table}. The condensation of the $m\Gamma_1^+$ irrep of this space group would correspond to the emergence of antiparallel magnetic moments with checkerboard configuration with mSG $I4/mcm.1$ (\#140.541). The Mn ions are at the Wyckoff positions $4c$ with site symmetry $4/m$, whose character table is shown in Table \ref{tab:4m_table}. The site symmetry irrep that induces $m\Gamma_1^+$ in the unit cell is $A_g$, i.e. $A_g\uparrow I4/mcm.1\rightarrow m\Gamma_1^+\oplus ...$, so for the collinear magnetic multipoles it is straightforward that the allowed ones are
\begin{equation}
    m_{\ell=1,3,5}=\left\{\begin{matrix}
        g_{xy(x^2-y^2)}m_z, & \textrm{even under } \rm{m}_{100}, \rm{m}_{110} \rightarrow m\Gamma_1^+,\\
        m_z,\; d_{z^2}m_z,\; g_{x^2y^2}m_z,\; g_{z^4}m_z, & \textrm{odd under } \rm{m}_{100}, \rm{m}_{110} \rightarrow m\Gamma_3^+.
\end{matrix}\right. 
\end{equation}

We have classified the multipoles depending on how they transform under the mirrors $\rm{m}_{100}$ and $\rm{m}_{110}$. As space group operations, these mirrors would be followed by a non-symmorphic translation connecting the Mn sites with opposite magnetic moment, either $(1/2,1/2,0)$ or $(0,0,1/2)$. As done before, we introduce operators in sublattice space that represent objects being in-phase, $\tau_0\in m\Gamma_1^+$, or out-of-phase, $\tau_z\in m\Gamma_3^+$, on the two sites. In this way the correct site multipoles are

\begin{equation}
    m\Gamma_1^+\; : \; m_{\ell=1,3,5}=\left\{\begin{matrix}
        g_{xy(x^2-y^2)} m_z\tau_0, \\
        \left\{1,\; d_{z^2},\; g_{z^4},\; g_{x^2y^2} \right\} m_z\tau_z.
\end{matrix}\right.
\end{equation}

We repeat the same operations for the in-plane magnetic multipoles, taking the product of $\left\{m_x, m_y\right\}$ with the $E_g$ electric quadrupoles and hexadecapoles. The product gives $E_g\times E_g = 2A_g \oplus 2B_g$ and focusing on the $A_g$ multipoles we obtain

\begin{equation}
    m\Gamma_1^+\; : \; m_{\ell=3,5}=\left\{\begin{matrix}
        \left\{d_{yz},\; g_{yz^3},\; g_{yz(y^2-3x^2)} \right\} m_x\tau_0 - \left\{d_{xz},\; g_{xz^3},\; g_{xz(x^2-3y^2)} \right\} m_y\tau_0, \\
        \left\{d_{xz},\; g_{xz^3},\; g_{xz(x^2-3y^2)} \right\} m_x\tau_z + \left\{d_{yz},\; g_{yz^3},\; g_{yz(y^2-3x^2)} \right\} m_y\tau_z.
\end{matrix}\right. 
\end{equation}

\begin{figure}[H]
    \centering
    \includegraphics[width=\linewidth]{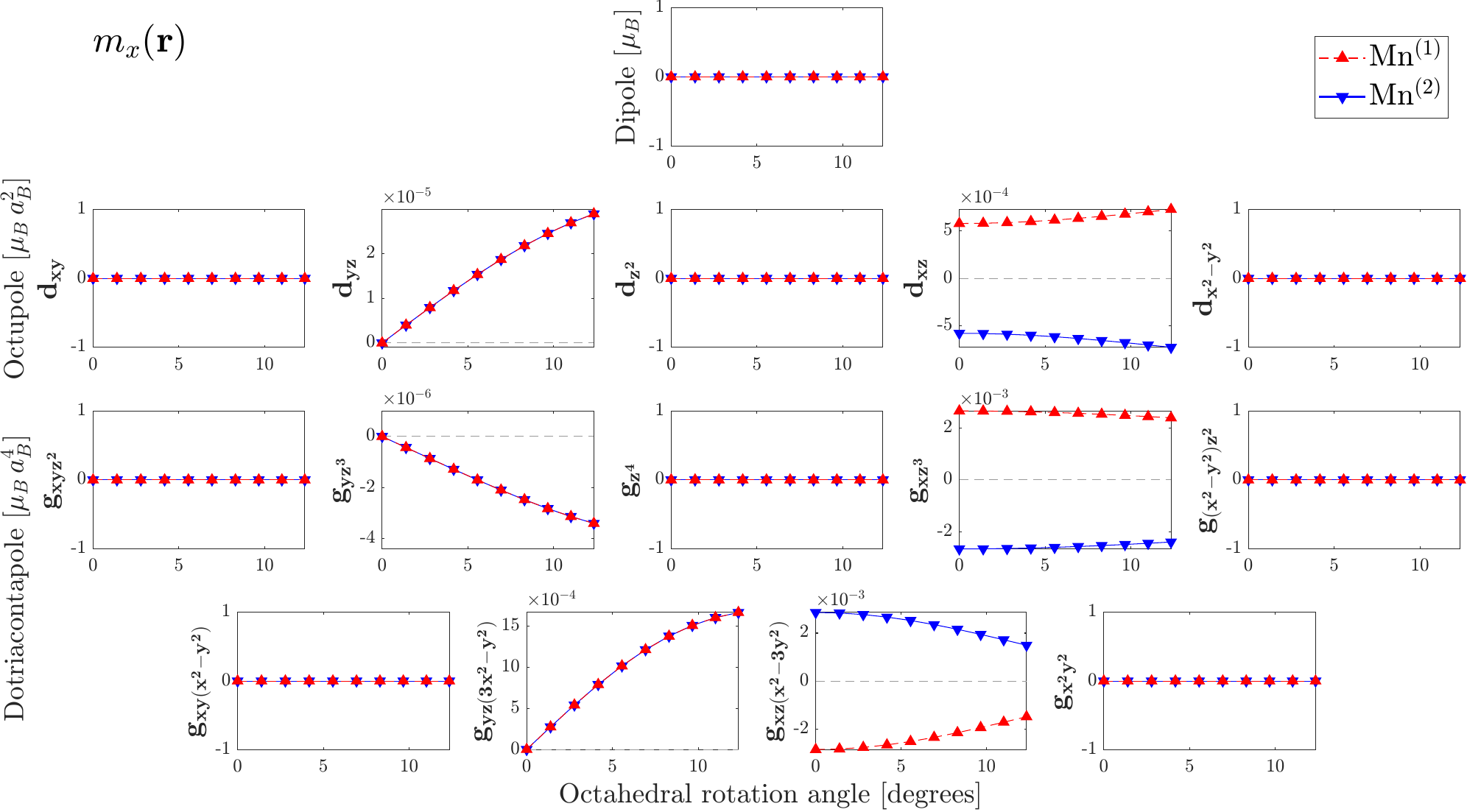}
    \caption{\label{fig:KMnF3_CubicHarmonics_SOx} Multipole coefficients $\alpha^{(1,2)}_{x,\ell m}$ obtained by projecting the $m_x(\mathbf{r})$ component of the spin density in KMnF$_3$ onto the inversion-even tesseral harmonics $T_{\ell m}$ up to $\ell=4$, as a function of the octahedral rotation angle.}
\end{figure}

\begin{figure}[H]
    \centering
    \includegraphics[width=\linewidth]{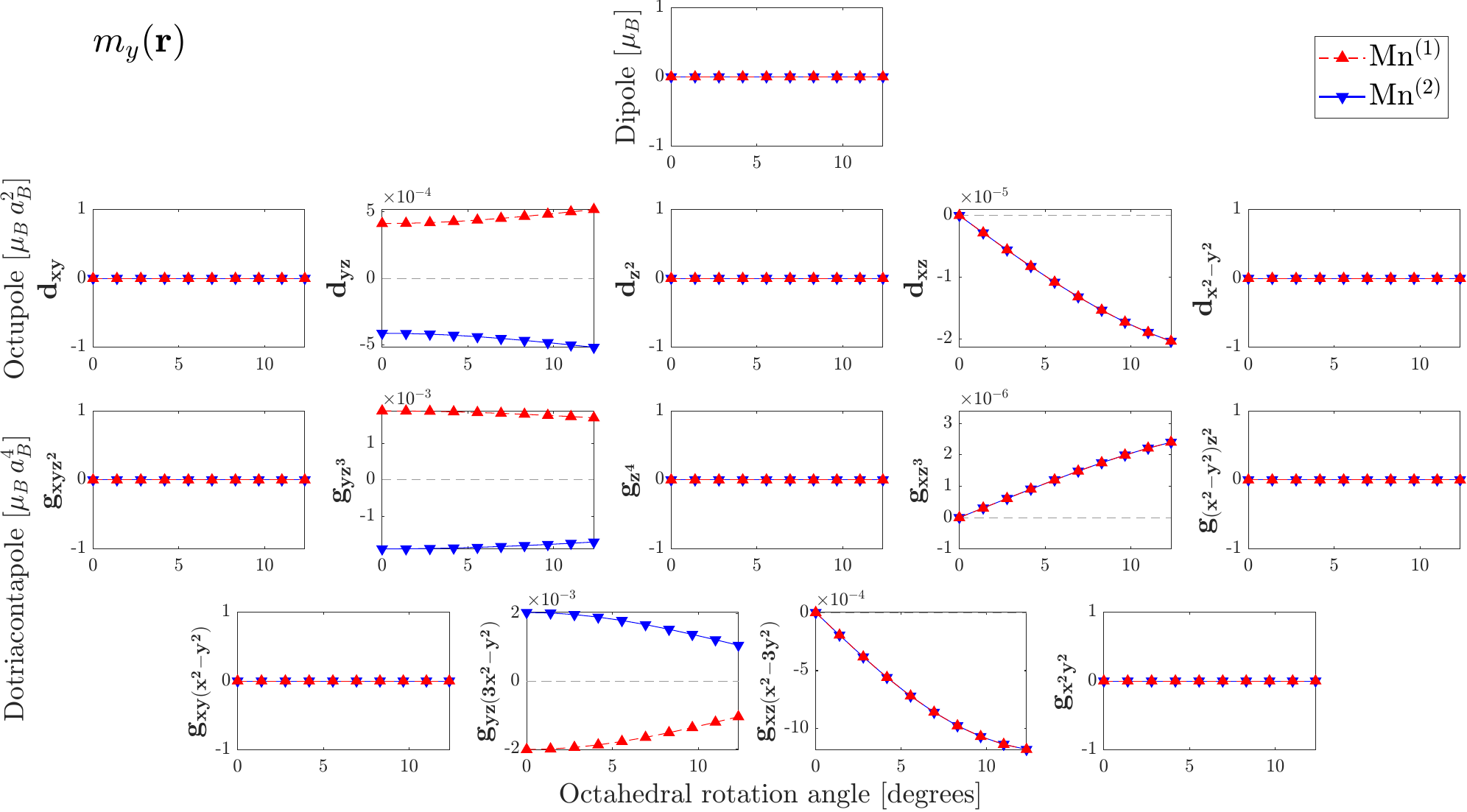}
    \caption{\label{fig:KMnF3_CubicHarmonics_SOy} Multipole coefficients $\alpha^{(1,2)}_{y,\ell m}$ obtained by projecting the $m_y(\mathbf{r})$ component of the spin density in KMnF$_3$ onto the inversion-even tesseral harmonics $T_{\ell m}$ up to $\ell=4$, as a function of the octahedral rotation angle.}
\end{figure}

\begin{figure}[H]
    \centering
    \includegraphics[width=\linewidth]{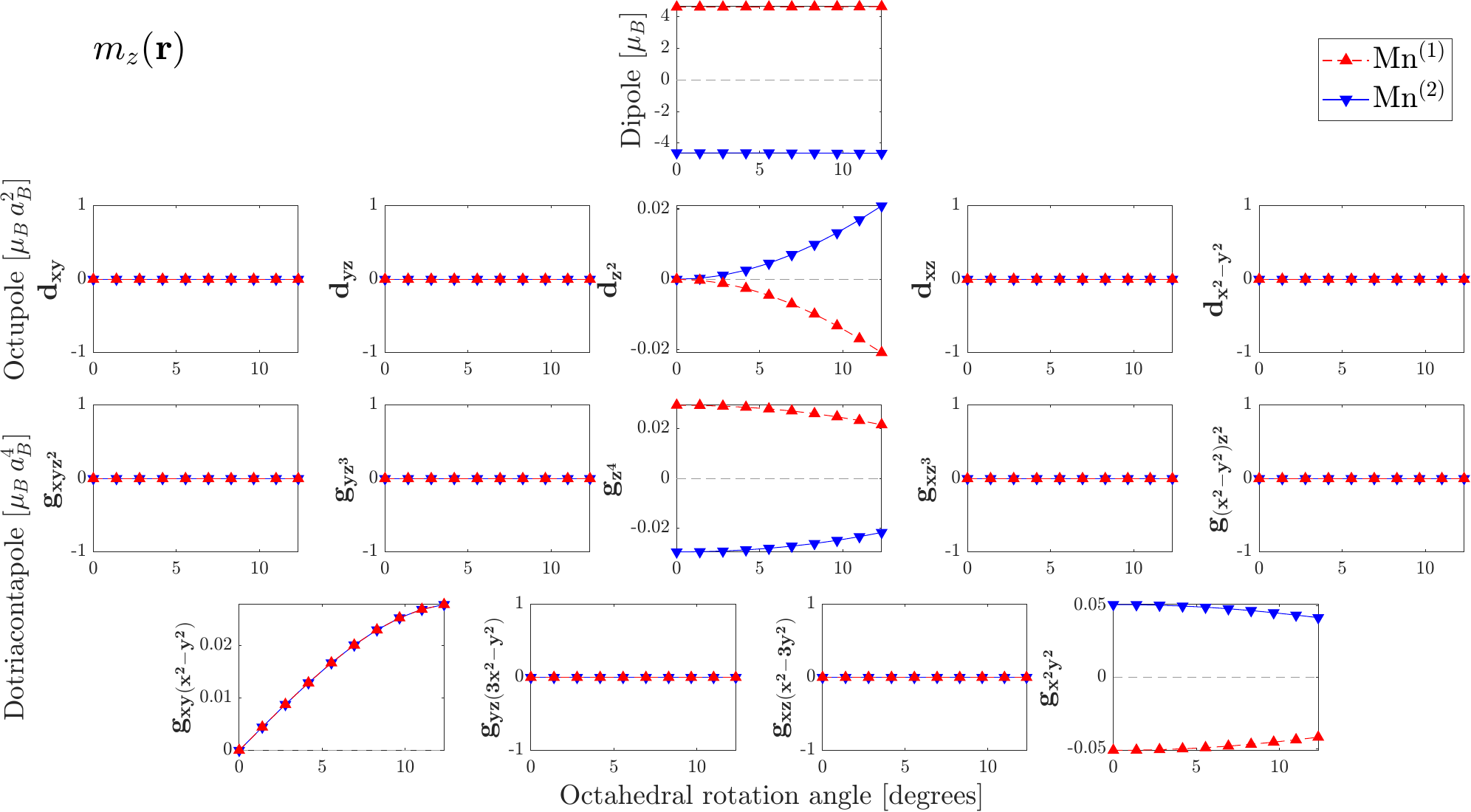}
    \caption{\label{fig:KMnF3_CubicHarmonics_SOz} Multipole coefficients $\alpha^{(1,2)}_{z,\ell m}$ obtained by projecting the $m_z(\mathbf{r})$ component of the spin density in KMnF$_3$ onto the inversion-even tesseral harmonics $T_{\ell m}$ up to $\ell=4$, as a function of the octahedral rotation angle.}
\end{figure}

This analysis is confirmed by our DFT calculations, shown in Figures \ref{fig:KMnF3_CubicHarmonics_SOx}, \ref{fig:KMnF3_CubicHarmonics_SOy} and \ref{fig:KMnF3_CubicHarmonics_SOz}. For the collinear $m_z$ magnetic multipoles, all of them are out-of-phase other than $g_{xy(x^2-y^2)}m_z$. For the noncollinear multipoles, the in-phase ones have opposite sign for $m_x$ and $m_y$, as expected from the previous equation. The lowest order multipole to be in-phase is $d_{yz}m_x-d_{xz}m_y$. Regarding piezomagnetism, this is in agreement with the expected selection rules for the piezomagnetic tensor for this mSG: no strain can lead to a weak ferromagnetic moment $m_z$, $\varepsilon_{yz}$ leads to a weak ferromagnetic moment $m_x$ and $\varepsilon_{xz}$ leads to a weak ferromagnetic moment $m_y$ and the two coefficients have same magnitude but opposite sign. These results can be generalized to any other tensor with Jahn symbol $aeV[V^2]$ and not just the piezomagnetic one.

\section{Landau Theory for Octahedral Rotation Induced Altermagnetism in KM\lowercase{n}F$_3$}

In perovskite KMnF$_3$, altermagnetism and zone-center multipoles emerge from the combination of zone-boundary antiferromagnetic order and the oxygen octahedral rotations, i.e. the antiferrodistortive order. This can be understood from a simple phenomenological model with a trilinear coupling between these three orders. In general, a trilinear term $\sim \gamma \cdot A\cdot B\cdot C$ between three degrees of freedom $A$, $B$, $C$  in the Landau free energy necessarily induces the order parameter $C$  when $A$ and $B$  are nonzero, because for nonzero $B$  and $C$  the minimum of free energy is at a nonzero value of $A$  independent of the sign of the coupling $\gamma$. Examples of this are observed in, for example, hybrid-improper ferroelectrics, and various charge density wave systems \cite{Benedek2011, Li2020, Ritz2023}. In KMnF$_3$, the same form of trilinear term exists between antiferomagnetic, altermagnetic, and oxygen octahedral rotation degrees of freedom, such that the combination of primary antiferromagnetic and structural orders induce an altermagnetic order. The magnitude of the altermagnetic multipole is linearly proportional to the magnitude of octahedral rotations, because for fixed antiferromagnetic order parameter magnitude, the coefficient of the first order term in the altermagnetic order is linearly proportional to the octahedral rotation order. 

In order to prove that there is a trilinear coupling in KMnF$_3$, we identify the relevant group representations and use standard group theory tools to predict which couplings are allowed \cite{Gallego2012, Isotropy}. The oxygen octahedral rotations in KMnF$_3$ are alternating in sign in every other layer of octahedra along the axis of rotations, and are denoted by the $-$ superscript in the Glazer notation. The irrep corresponding to this structural distortion is the 3D $R_5^-$ of the parent space group $Pm\bar{3}m$. The low temperature $I4/mcm$  structure is obtained when octahedra rotates around the $c$  axis, which corresponds to the  $R_5^-(0,0,a)$  direction of the order parameter. The G-type antiferromagnetic order in a cubic perovskite can be shown to transfrom as the $mR_5^-$, which is coincidentally the same irrep as the structural distortion in KMnF$_3$, apart from being time-reversal odd. This can be understood from the fact that under crystallographic symmetry transformations, angular momentum behaves the same way as as a static rigid rotation. For magnetic moments along the crystallographic $c$  axis, which is observed in KMnF$_3$, the relevant direction of the antiferromagnetic order is $mR_5^-(0,0,a)$. 

The product of these two irreps give 
\begin{equation}
R_5^-\otimes mR_5^- = m\Gamma_1^+\oplus m\Gamma_3^+\oplus m\Gamma_4^+\oplus m\Gamma_5^+
\end{equation}
which means that there are trilinear terms between the octahedral rotations, the G-type antiferromagnetic order, and order parameters transform as the irreps on the right hand side. All these irreps are time-reversal odd and inversion even zone center representations, which means that they are either ferromagnetic ($m\Gamma_4^+$) or altermagnetic ($m\Gamma_1^+$, $m\Gamma_3^+$, $m\Gamma_5^+$) \cite{Fernandes2024Topological}. For a specific direction of the octahedral rotations and the antiferromagnetic moments, one or more of the irreps on the right hand side are relevant. In particular, the combination relevant to KMnF$_3$ gives $R_5^-(0,0,a)\otimes mR_5^-(0,0,b) = m\Gamma_1^+(c)$, so the only zone-center irrep induced by the combination of these two primary orders is the  $m\Gamma_1^+$ altermagnetic order. In other words, if we denote the antiferromagnetic $mR_5^-(0,0,b)$ as $L$, the octahedral rotations as $R$, and the $m\Gamma_1^+$  altermagnetic order as $A$, the Landau free energy $\mathcal{F}$ for $A$  has the form 
\begin{equation}
\mathcal{F}=\frac{\alpha}{2} A^2 +\gamma ARL
\end{equation}
which is minimum for $A=-\frac{\gamma}{\alpha}RL$.

$m\Gamma_1^+$ in the cubic group gives rise to a 32-pole \cite{Fernandes2024Topological} with spin density 
\begin{equation}
    \textbf{m}(\textbf{r})=yz(y^2-z^2)m_x + zx(z^2-x^2)m_y +xy(x^2-y^2)m_z  . 
\end{equation}
In the absence of SOC, only the collinear part that depends on $m_z$ is nonzero, and is linear in the octahedral rotation angle as shown in the main text and Supp. Fig.~\ref{fig:KMnF3_CubicHarmonics_SOz}. The in-plane components corresponding to this multipole are also nonzero in the presence of SOC, and are also linear in the octahedral rotation angle (for small rotations) as shown in Fig.~\ref{fig:KMnF3_CubicHarmonics_SOx}  and \ref{fig:KMnF3_CubicHarmonics_SOy}, where they are decomposed in the in-phase multipoles $g_{yz(3x^2-y^2)}m_x$, $g_{yz^3}m_x$ and $g_{xz(x^2-3y^2)}m_y$, $g_{xz^3}m_y$.

\section{One-orbital toy model}
\label{sec:toy}

It is instructive to consider a simple toy model that captures the essential physics of how the SOC modifies spin densities and hence multipolar patterns. The aim of this model is twofold: (i) to demonstrate explicitly why noncollinear multipoles appear already at linear order in the SOC strength $\lambda$, whereas collinear components are only modified at quadratic order; and (ii) to show how higher-order multipolar textures, such as a 32-pole, can emerge even in a system built from $d$-orbitals alone.  

To this end, we construct the simplest possible starting point: a single $d$-orbital state with spin polarization along $z$, on which we subsequently act with the SOC perturbation. Despite its minimal character, this model already illustrates the key symmetry arguments and angular-momentum selection rules that underlie the more general results presented in the main text.

Let us start from a state $\ket{\psi_0}=\ket{2,0}\ket{\uparrow}$ that is a tensor product of eigenstates of the angular momentum $L^2$ and the spin $S^2$ operators , such that

\begin{align}
    L^2\ket{l,m} &= l(l+1)\ket{l,m}, \\
    L_z\ket{l,m} &= m, \\
    S^2\ket{\uparrow} &= \frac{1}{2}\left(\frac{1}{2}+1\right)\ket{\uparrow}, \\
    S_z\ket{\uparrow} &= \frac{1}{2}\ket{\uparrow}.
\end{align}

The density matrix for this state has only one component, which we can represent in real space as
\begin{align}
\begin{split}
    \bra{\mathbf{r}}\hat{\rho}_0\ket{\mathbf{r}} &= \braket{\mathbf{r}|\psi_0}\braket{\psi_0|\mathbf{r}} \\
    &= \braket{\mathbf{r}|2,0}\braket{2,0|\mathbf{r}}\ket{\uparrow}\bra{\uparrow} \\
    &= |Y_{2,0}(\mathbf{r})|^2\ket{\uparrow}\bra{\uparrow},
\end{split}
\end{align}
where we denote with $Y_{l,m}$ the spherical harmonics times an appropriate radial function that makes them a complete (orthonormal) set in 3D. For the purpose of this toy calculation we choose to ignore the radial part since we are mainly interested in the angular behavior. The state $\psi_0$ will therefore give rise to a charge density and a spin-density in the $z$-direction that are identical to each other and behave like $\sim Y_{4,0}$ given the contraction rules of the spherical harmonics.

\begin{figure}[]
    \centering
    \includegraphics[width=0.99\linewidth]{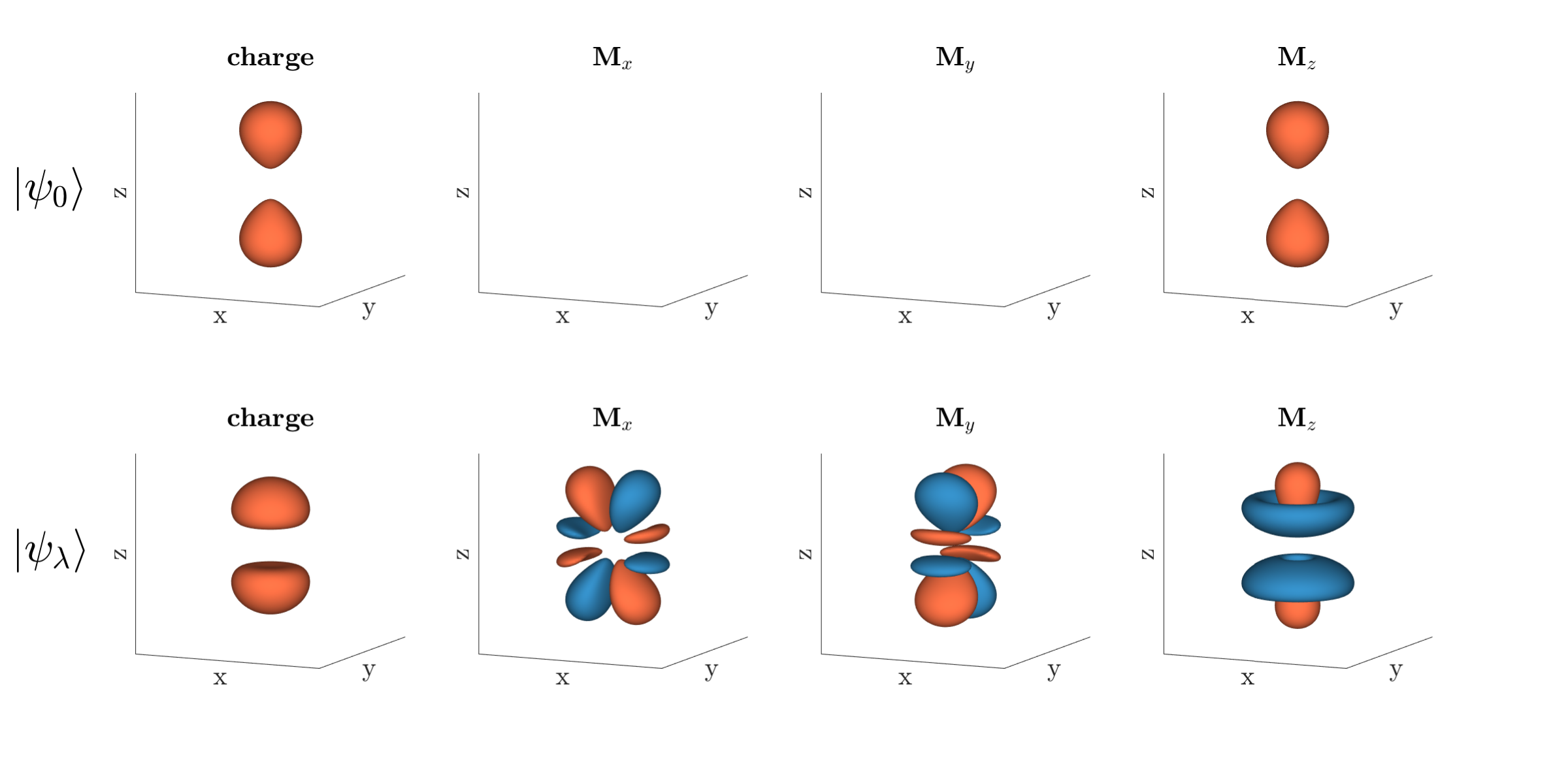}
    \caption{Charge and spin densities for the one-orbital toy model. In the first row, the state $\ket{\psi_0}$ without spin-orbit coupling; in the second row, the state $\ket{\psi_1}$ with the spin-orbit coupling introduced at second order in perturbation theory.}
    \label{fig:toy1orb}
\end{figure}

Now we introduce a perturbation to the state in the form of a spin-orbit coupling $\lambda\mathbf{L}\cdot\mathbf{S}$ which, given the state we have chosen, reduces to just $-\frac{\lambda}{2}L_+S_-$. Thus, up to second order in $\lambda$ we have a perturbed state

\begin{align}
\begin{split}
    \ket{\psi_\lambda} &= \ket{\psi_0} + \lambda\ket{\delta\psi} + \lambda^2\ket{\delta^2\psi} + O(\lambda^3),\\
    \ket{\delta\psi} &= -\frac{\sqrt{6}}{2} \ket{2,1}\ket{\downarrow}, \\
    \ket{\delta^2\psi} &= \frac{\sqrt{6}}{4} \ket{2,1}\ket{\downarrow} - \frac{3}{2} \ket{2,0}\ket{\uparrow}.
\end{split}
\end{align}

The density matrix operator for the perturbed state $\ket{\psi_\lambda}$ now has also off diagonal elements

\begin{align}
\begin{split}
    \hat{\rho}(\mathbf{r}) = &(1-3\lambda^2)\hat{\Pi}_0(\mathbf{r}) \\
    &+ (- \lambda\frac{\sqrt{6}}{2} +  \lambda^2\frac{\sqrt{6}}{4}) \braket{\mathbf{r}|2,1}\braket{2,0|\mathbf{r}}\ket{\downarrow}\bra{\uparrow} \\
    &+ (- \lambda\frac{\sqrt{6}}{2} +  \lambda^2\frac{\sqrt{6}}{4}) \braket{\mathbf{r}|2,0}\braket{2,1|\mathbf{r}}\ket{\uparrow}\bra{\downarrow} \\
    &+ \lambda^2\frac{3}{2} \braket{\mathbf{r}|2,1}\braket{2,1|\mathbf{r}}\ket{\downarrow}\bra{\downarrow},
\end{split}
\end{align}

\begin{equation}
    \hat{\rho}_{ij}(\mathbf{r}) = \begin{pmatrix}
(1-3\lambda^2)|Y_{2,0}(\mathbf{r})|^2 & \frac{\sqrt{6}}{2}(\frac{\lambda^2}{2}-\lambda)Y_{2,0}(\mathbf{r})Y^*_{2,1}(\mathbf{r}) \\
\frac{\sqrt{6}}{2}(\frac{\lambda^2}{2}-\lambda)Y^*_{2,0}(\mathbf{r})Y_{2,1}(\mathbf{r}) & \frac{3}{2}\lambda^2|Y_{2,1}(\mathbf{r})|^2, \\
\end{pmatrix}
\end{equation}
where we disregarded the terms with powers higher than $\lambda^2$. The components of the spin density along the cartesian directions are 
\begin{equation}
    m_x(\mathbf{r}) = \frac{\sqrt{6}}{4}(\frac{\lambda^2}{2}-\lambda) \biggl( Y_{2,0}(\mathbf{r})Y^*_{2,1}(\mathbf{r}) + Y^*_{2,0}(\mathbf{r})Y_{2,1}(\mathbf{r}) \biggr)
\end{equation}
\begin{equation}
    m_y(\mathbf{r}) = i\frac{\sqrt{6}}{4}(\frac{\lambda^2}{2}-\lambda) \biggl( -Y_{2,0}(\mathbf{r})Y^*_{2,1}(\mathbf{r}) + Y^*_{2,0}(\mathbf{r})Y_{2,1}(\mathbf{r}) \biggr)
\end{equation}
\begin{equation}
    m_z(\mathbf{r}) = (1-3\lambda^2)|Y_{2,0}(\mathbf{r})|^2 + \frac{3}{2}\lambda^2|Y_{2,1}(\mathbf{r})|^2,
\end{equation}
where the angular behavior of the noncollinear components is $\sim Y_{4,\pm1}(\mathbf{r})$ and is linear in the SOC constant $\lambda$. On the other hand, the collinear component is modified at quadratic order in $\lambda$.

\bibliography{MAIN}